\documentclass[prd, onecolumn, nofootinbib, floatfix]{revtex4}

\usepackage{amsmath}
\usepackage{graphicx}
\usepackage{dcolumn}
\usepackage{bm}
\usepackage{epsfig}
\usepackage{amssymb,latexsym,mathrsfs}
\usepackage{graphicx}
\usepackage{color}
\usepackage{hyperref}

\hypersetup{
    colorlinks=true,
    linkcolor=red,
    citecolor=blue,
} 

\newcommand{\be}{\begin{equation}}
\newcommand{\ee}{\end{equation}}
\newcommand{\bs}{\begin{split}} 
\newcommand{\bea}{\begin{eqnarray}}
\newcommand{\eea}{\end{eqnarray}}

\newcommand{\geff}{G_{\rm eff}}
\newcommand{\Mpl}{M_{\rm pl}}
\newcommand{\kap}{\kappa_}

\begin{document}

\title{Fab 5: Noncanonical Kinetic Gravity, Self Tuning, and Cosmic Acceleration} 
\author{Stephen A.\ Appleby$^1$, Antonio De Felice$^2$, Eric V.\ 
Linder$^{1,3}$} 
\affiliation{$^1$ Institute for the Early Universe WCU, Ewha Womans 
University, Seoul, Korea} 
\affiliation{$^2$ ThEP's CRL, NEP, The Institute for Fundamental Study, 
Naresuan University, Phitsanulok 65000, Thailand}
\affiliation{$^3$ Berkeley Lab \& University of California, Berkeley, 
CA 94720, USA}

\begin{abstract}
We investigate circumstances under which one can generalize Horndeski's 
most general scalar-tensor theory of gravity.  Specifically we demonstrate 
that a nonlinear combination of purely kinetic gravity terms can give rise 
to an accelerating universe without the addition of extra propagating 
degrees of freedom on cosmological backgrounds, and exhibit self tuning to 
bring a large cosmological 
constant under control.  This nonlinear approach leads to new properties 
that may be instructive for exploring the behaviors of gravity. 
\end{abstract}

\date{\today} 

\maketitle

\section{Introduction} 

Theories of gravitation on cosmic scales have become an area of intense 
interest, both as a possible explanation for the observed cosmic 
acceleration and as an exploration of consistent extensions of general 
relativity.  Generically such extensions lead to additional degrees of 
freedom, e.g.\ scalar modes in scalar-tensor theories, with 
possible pitfalls of higher than second order derivative field equations 
that may lack a well posed initial value formulation, or of ghosts 
and other instabilities. 

Horndeski in 1974 wrote the most general scalar-tensor theory giving 
second order field equations in four dimensional spacetimes \cite{horndeski}. 
In an alternate view, Galileon theories \cite{nicolis,deffayet1,deffayet2}, 
shift symmetric scalar fields possessing nonlinear combinations 
of field derivatives, 
have recently been 
studied with interest as sound models capable of cosmic acceleration. 
Also recently, as an approach to solve the cosmological constant problem, 
a linear combination of four 
terms called the Fab Four has been identified \cite{fab4,fab4long,fab4new} 
as the unique terms allowing self tuning vacua that can cancel 
a large bare $\Lambda$ term. 

Here we draw on aspects of all three of these approaches to 
demonstrate that nonlinear combinations of terms involving shift symmetric 
scalar fields can possess interesting advantages 
and properties.  This extension of the ``most general'' 
scalar-tensor theory retains second order field equations and avoids 
pathologies on symmetric spacetimes such as the usual 
Friedmann-Lema\^\i tre-Robertson-Walker (FLRW) and de Sitter cosmologies.  
While the nonlinear approach can be applied quite generally, 
we give a proof of principle using a simple 
example of purely kinetic couplings with noncanonical forms, extending 
the ``purely kinetic gravity'' of \cite{gub}.  

Our example employs a nonlinear combination of 
the standard kinetic term and derivative coupling to the Einstein tensor.  
Besides possessing second order field equations it does not add any further 
propagating degrees of freedom, and it can achieve lasting cosmic 
acceleration unlike the linear, canonical, purely kinetic gravity theory of 
\cite{gub}, avoid at least some instabilities unlike the derivatively 
coupled Galileon theory investigated by \cite{applin}, and self tune away 
a cosmological constant like the Fab Four. 

Deeper implications exist beyond our simple proof of principle.  
We emphasize that the example given is intended purely as a proof of 
principle to inspire further investigation into the theoretical properties 
of general nonlinear combinations, and not as a fit to observations. 

In Sec.~\ref{sec:method} we explain our nonlinear generalization 
procedure and the conditions under which no additional propagating 
degrees of freedom are generated.  The equations of motion are solved 
in Sec.~\ref{sec:eom} on a FLRW background, giving 
the cosmic and field evolution complete with attractors, revealing 
two distinct ways of approaching a de Sitter asymptotic state.  
Section~\ref{sec:tune} demonstrates the self tuning properties of the 
theory, erasing an initial cosmological constant.  The perturbed equations in 
Sec.~\ref{sec:perturb} yield the no-ghost and stability conditions and 
the evolution of the effective Newton's constant $\geff$.  We discuss 
various implications of the results in Sec.~\ref{sec:concl}.

\section{Promotion to Nonlinear Function} \label{sec:method} 

The Einstein-Hilbert action of general relativity is extremely simple, 
condensing all the gravitational influence into the Ricci scalar 
curvature $R$.  To allow cosmic acceleration, however, one must add 
either a cosmological constant $\Lambda$ or additional degrees of freedom 
such as 
a scalar field $\phi$, e.g.\ with a potential and canonical kinetic term 
$X\equiv (-1/2)g^{\mu\nu}\phi_\mu \phi_\nu$ where $\phi_\mu=\nabla_\mu \phi$.  
The cosmological constant, or the field potential, raises issues of fine 
tuning and naturalness: why don't high energy radiative corrections 
affect the form and magnitude to something characteristic of the early 
universe?  We therefore do not employ either (except in Sec.~\ref{sec:tune} 
where we erase them). 

A canonical kinetic term cannot by itself give rise to acceleration but 
noncanonical (but still minimally coupled) kinetic terms can, called 
k-essence \cite{armend1,armend2,chiba}, or in the absence of any potential, 
purely kinetic k-essence (e.g.\ \cite{07050400}).  One can think of this as 
promoting the Lagrangian term linear in the canonical kinetic contribution 
to a function.  This generically gives an extra degree of freedom, in 
that the sound speed is no longer fixed to the speed of light.  

Alternately, one could promote the Ricci scalar term to a function, 
hence $f(R)$ theories \cite{fRreview}.  This again adds a degree of freedom 
and one can 
view these as coupled scalar-tensor theories.  Similarly one can have 
theories involving the Gauss-Bonnet combination 
$G_{\rm GB}=R^{abcd} R_{abcd}-4R^{ab} R_{ab} +R^2$ of the Riemann tensor, 
Ricci tensor, and Ricci scalar, either linearly in $G_{\rm GB}$ or promoted 
to a function \cite{gb,gbfn}.  In generalized Galileon theories one can 
promote the coefficients of the standard Galileon terms to functions of 
the canonical kinetic term, for example \cite{gengal1,gengal2}. 

Taking these examples as motivation, suppose we take the Horndeski theory 
action, composed of the linear 
combination of several terms, and instead promote them to nonlinear 
functions or combinations.  In generality we cannot do this without 
resulting in higher than second order equations of motion or adding 
unconstrained degrees of freedom.  However, in specific circumstances 
we can.  For example a theory involving a function of the Ricci scalar 
and Gauss-Bonnet term $f(R,G_{GB})$ can be sound \cite{frgb}.  In 
Fab Four terminology, this mixes George and Ringo (though we do not allow 
the field potentials).  This is permissible because of particular 
symmetries within these terms. 

In purely kinetic gravity theories, similar symmetries impose a unique 
Lagrangian involving only the Einstein tensor coupled to the field 
derivatives \cite{amendola} (which Fab Four term John basically replicates).  
Up to mass dimension 6, the action is just 
the linear combination of the canonical kinetic term and the Einstein 
coupled kinetic term \cite{gub}, effectively giving a disformal field theory. 
This could achieve transient cosmic acceleration but not an asymptotic 
de Sitter state, and was later shown to have ghosts \cite{applin}. 
Allowing for an arbitrary constant coefficient of the canonical kinetic 
term, one could achieve a de Sitter asymptote but the theory has an 
early time Laplace instability \cite{applin}. 

Merging these two approaches of nonlinear function promotion and 
purely kinetic terms of great simplicity, we examine as a specific 
example nonlinear functions of the canonical and the Einstein coupled 
kinetic terms.  The combination of nonlinearity and noncanonical nature 
delivers new characteristics to the theory.  Since this ``hip-hop'' kinetic 
evolution extends the Fab Four self tuning possibilities, among other 
properties, we call this new Lagrangian term Fab 5 Freddy.  As the line 
``Fab 5 Freddy told me everybody's fly'' from Blondie's {\it Rapture\/} 
\cite{rapture} predicts, this term also enables cosmic acceleration and 
an asymptotic de Sitter behavior, indeed in multiple ways. 

The action we study in detail is 
\begin{equation}
S=\int d^{4}x\sqrt{-g}\left[\frac{M_{{\rm pl}}^{2}}{2}\, R+c_{1}X+f\left(c_{2}X+{c_{G} \over M^{2}}G^{\mu\nu}\phi_{\mu}\phi_{\nu}\right)\right]+S_{{\rm m}}\,,\label{eq:act1}
\end{equation}
where $G^{\mu\nu}$ is the Einstein tensor associated to the metric 
$g_{\mu\nu}$, $S_{{\rm m}}$ represents the action for the matter fields, and 
$M$ is a mass scale to keep $c_G$ dimensionless, where we normalize to
$M = H_{0}$.  
When the function $f$ is linear then this is the 
derivatively coupled Galileon (using only ${\mathcal L}_2$ in \cite{applin}), 
generalizing the purely kinetic gravity model by allowing a free constant 
coefficient for the canonical term.  

To study the effects of the nonlinear promotion we consider two cases: 
1) $c_1=0$, so the canonical and Einstein coupled kinetic terms are 
directly coupled nonlinearly, and 2) $c_2=0$, so only the derivative 
coupling appears nonlinearly.  This allows us to compare these two 
different theories with the same linear limit. 

We can rewrite the action in terms of a Lagrange multiplier field $\chi$, as 
\begin{equation}
S=\int d^{4}x\sqrt{-g}\left[\frac{M_{{\rm pl}}^{2}}{2}\, R+c_{1}X+f(\chi)+\left(c_{2}X+{c_{G} \over M^{2}} G^{\mu\nu}\phi_{\mu}\phi_{\nu}-\chi\right)\,\frac{df}{d\chi}\right]+S_{{\rm m}}\,.\label{eq:act2}
\end{equation}
Varying the action (\ref{eq:act2}) in terms of $\chi$ we find 
\begin{equation}
\left(c_{2}X+{c_{G} \over M^{2}}G^{\mu\nu}\phi_{\mu}\phi_{\nu}-\chi\right)\,\frac{d^{2}f}{d\chi^{2}}=0\,. 
\end{equation} 
This has the solution 
\begin{equation}
\chi=c_{2}X+{c_{G} \over M^{2}}G^{\mu\nu}\phi_{\mu}\phi_{\nu}\ ,\label{eq:lagrmul}
\end{equation} 
except at particular points for which $f_{\chi\chi}=0$ (and note that in 
the linear case $\chi$ is moot).  Subscripts $\chi$ denote derivatives 
with respect to $\chi$.  
By re-inserting the solution Eq.~(\ref{eq:lagrmul}) back into 
Eq.~(\ref{eq:act2}), we verify that we obtain the original action 
Eq.~(\ref{eq:act1}). 

Introducing a Lagrange multiplier field $\chi$ helps understanding of the 
independent degrees of freedom. In particular, such a Lagrange multiplier 
can be coupled with other elements (such as the Ricci scalar, as in the 
$f(R)$ theories, or the Einstein tensor, as in our case). Both the Ricci 
scalar and the Einstein tensor are functions of a second derivative of the 
metric. Therefore by integrating by parts, a time-derivative for the 
Lagrange multiplier may appear. In this case, such a Lagrange multiplier 
can in general acquire a kinetic term, and it may start propagating. This 
situation, as already said, is common to those theories which can be written 
in terms of a Lagrange multiplier coupled to a second-order operator, e.g.\ 
as in $f(R)$ or $R+f(G_{\rm GB})$. The theory $f(R,G_{\rm GB})$ introduces 
two Lagrange multipliers. This theory is quite interesting as it has been 
proven that only one of these two new scalar degrees of freedom will 
propagate on Friedmann-Lema\^\i tre-Robertson-Walker backgrounds 
\cite{frgb}. On the other hand, both these degrees of freedom do propagate 
on anisotropic backgrounds. Therefore whether or not these Lagrange 
multipliers propagate or not depends on the chosen theory.  

We will see that the theory at hand, Eq.~(\ref{eq:act1}), will not 
introduce on cosmological backgrounds any new degree of freedom.  However, 
we will find that the high-$k$ limit (where $k$ is the wavemode) dispersion 
relation of perturbations will be modified, leading to a scale-dependent 
speed of propagation, i.e.\ $c_s^2\propto k^2$. This is indeed similar to 
what happens in the $f(R,G_{\rm GB})$ case.  On a formal level it will be 
interesting to study eventually our theory on anisotropic backgrounds to 
see whether or not the Lagrange multiplier will start propagating and we 
will discuss this issue in a future project.

\section{Equations of Motion and Evolution} \label{sec:eom} 

We give the general covariant background equations of motions in 
Appendix~\ref{sec:coveom}.  Here we specialize to a homogeneous and 
isotropic spacetime where the metric is FLRW.  The theory then has the 
property that the equations of motion for the action remain second order. 
We include a barotropic fluid (i.e.\ matter and radiation) with energy 
density $\rho$ and pressure $P$ and assume spatial flatness for simplicity.  
The background equations of motion for the action of Eq.~(\ref{eq:act2}) 
are then 
\begin{eqnarray}
3\Mpl^{2}\,{H}^{2}&=&\rho+\frac{1}{2}c_{1}\dot{\phi}^{2}+ 
\frac{1}{2}c_{2}f_\chi \dot{\phi}^{2}+f_\chi\,\chi-f+ 
9{c_{G} \over M^{2}} f_\chi {H}^{2}\dot{\phi}^{2}\ , \label{eq:fried1}\\
2\left(\Mpl^{2}- {c_G \over M^{2}} f_\chi {\dot{\phi}}^{2}\right)\dot{H}&=&-P-3\,\Mpl^{2}{H}^{2}- 
\frac{1}{2}c_{{1}}{\dot{\phi}}^{2}-f \label{eq:fried2}\\ 
&&-\left(\frac{1}{2}c_{{2}}{\dot{\phi}}^{2}-3\, {c_{G} \over M^{2}}{H}^{2}{\dot{\phi}}^{2}-\chi-4\, {c_{G} \over M^{2}}H\dot{\phi}\ddot{\phi}\right)f_\chi+ 
2 {c_{G} \over M^{2}}H{\dot{\phi}}^{2}\dot{f_\chi} \ ,\nonumber\\ 
\left(c_{1}+c_{2}f_\chi+6 {c_{G} \over M^{2}}{H}^{2}f_\chi\right)\,\ddot{\phi}&=& 
-3c_{1}H\dot{\phi}-12{c_{G} \over M^{2}} f_\chi\,\dot{\phi} H\dot{H}-
(\dot{f_\chi}+3 Hf_\chi)\left(c_{2}+6 {c_{G} \over M^{2}}{H}^{2}\right)\dot{\phi} \ ,\\ 
\chi&=& \left(\frac{1}{2}c_{2}+3{c_{G} \over M^{2}} H^{2}\right)\dot{\phi}^{2}\ ,\label{eq:eomchi}\\ 
\dot{\rho}&=&-3H\,(\rho+P) \ ,  
\end{eqnarray} 
where $H=\dot a/a$ is the Hubble expansion rate of the scale factor $a$. 
 
Linear perturbations about the background are important for calculating the 
growth of structure, which we consider in Sec.~\ref{sec:perturb}, but also 
for analyzing the degrees of freedom.  Details of the equations are given 
in Appendix~\ref{sec:apxpert} but here we note a key point.  The coupled 
system of perturbed equations for the two metric potentials, the barotropic 
fluid density and velocity, the $\phi$ scalar field, and the Lagrange 
multiplier scalar field $\chi$ does not possess any time derivatives 
$\dot{\delta\chi}$.  This indicates that $\chi$ is merely an auxiliary field 
with no dynamics but rather an algebraic constraint, and is uniquely 
determined by the other fields.  This arises because the Einstein tensor 
within $f(\chi)$ only depends on first derivatives and not second derivatives 
in the Robertson-Walker background, i.e.\ only $H^2$ appears.  

To obtain the solutions to the evolution of the expansion $H$ and field 
$\phi$, we put the background equations in the form of an autonomous system 
of coupled equations, using the dimensionless parameters 
$\bar{H} \equiv H/H_{0}$ and $x \equiv \phi'/M_{\rm pl}$, where primes 
denote derivatives with respect to $N = \ln a$.  Then 
\begin{eqnarray} 
\label{eq:67} & &  x' = {\lambda \gamma - \omega \alpha  \over \alpha\sigma - \lambda\beta} \\ \label{eq:68} & & \bar{H}' = -{\gamma \over \alpha} - {\beta \over \alpha} x' 
\end{eqnarray} 
where 
\begin{eqnarray} 
& & \alpha =  2\bar{H} -6 f_{\chi} c_{\rm G} \bar{H}^{3}x^{2} -2 c_{\rm G} f_{\chi\chi} \bar{H}^{4} x^{2} \left( c_{2} \bar{H} x^{2} + 12c_{\rm G} \bar{H}^{3}x^{2} \right) \\ & & \beta =  -2c_{\rm G} f_{\chi\chi}\bar{H}^{4}x^{2} \left( c_{2} \bar{H}^{2}x + 6 c_{\rm G} \bar{H}^{4}x \right) - 4 f_{\chi} c_{\rm G}\bar{H}^{4} x \\ 
& & \gamma = 3\bar{H}^{2} +f_{\chi}\left( {c_{2} \over 2} \bar{H}^{2} x^{2}  - \chi - 3c_{\rm G}  \bar{H}^{4} x^{2} \right) + {\Omega_{r0} \over a^{4}} + f + {c_{1} \over 2} \bar{H}^{2}x^{2}  \\  
& & \sigma =  c_{2} f_{\chi} \bar{H}^{2} + 6 f_{\chi}  c_{\rm G} \bar{H}^{4} + f_{\chi\chi}  \bar{H}^{2} x  \left( c_{2} + 6 c_{\rm G} \bar{H}^{2} \right)\left( c_{2} \bar{H}^{2} x +6 c_{\rm G} \bar{H}^{4}x \right) + c_{1}\bar{H}^{2}\\ 
& & \lambda = f_{\chi} c_{2} \bar{H}x + 18f_{\chi} c_{\rm G} \bar{H}^{3} x +  f_{\chi \chi} \bar{H}^{2} x \left( c_{2} + 6 c_{\rm G} \bar{H}^{2}\right) \left( c_{2} \bar{H} x^{2} + 12 c_{\rm G} \bar{H}^{3} x^{2}\right) + c_{1} \bar{H} x  \\ 
& & \omega = 3f_{\chi} \bar{H}^{2}x \left( c_{2} + 6 c_{\rm G} \bar{H}^{2} \right) + 3 c_{1}\bar{H}^{2} x 
\end{eqnarray} 
with $\Omega_{r0}$ the dimensionless radiation energy density today and 
\begin{equation} 
\chi = {c_{2} \over 2} \bar{H}^{2} x^{2} + 3c_{\rm G} \bar{H}^{4} x^{2} \ . 
\label{eq:chi} 
\end{equation} 

To ensure the accuracy of our numerical solution we use as a check the 
constraint equation~(\ref{eq:fried1}), written in the dimensionless 
parameters as 
\begin{equation} 
\label{eq:op1} \bar{H}^{2} = {\Omega_{m0} \over a^{3}} + {\Omega_{\rm r 0} \over a^{4}} + {1 \over 3} \left( f_{\chi} \chi - f + 9f_{\chi} c_{\rm G} \bar{H}^{4} x^{2} + f_{\chi} {c_{2} \over 2} \bar{H}^{2} x^{2}  + {c_{1} \over 2} \bar{H}^{2} x^{2} \right) \ , 
\end{equation} 
with $\Omega_{m0}$ the dimensionless matter density today.  
The quantity in parentheses can be viewed as an effective dark energy density. 
An effective dark energy pressure can similarly be defined using 
Eq.~(\ref{eq:fried2}), with the effective dark energy equation of state 
ratio $w_\phi=P_\phi/\rho_\phi$.

\subsection{Early Time Evolution} \label{sec:early} 

As in \cite{applin}, one can identify the early and late time asymptotic 
solutions.  At early times, during radiation or matter domination, when 
$\bar H^2\gg1$, one generally has $\chi\approx 3c_G \bar H^4x^2$ and 
$\Omega_\phi\ll1$ (if one fine tunes the $c_1$ or $c_2$ terms to dominate 
instead then the energy density would decay as $\rho_\phi\sim a^{-6}$ and 
hence be uninteresting).  In this case the solution becomes 
\bea 
x&\sim&a^{3[1+3w_b+4e(1+w_b)]/[2(1+2e)]} \\ 
\chi&\sim &a^{-3(1-w_b)/(1+2e)} \ , 
\eea 
where $w_b$ is the barotropic equation of state (0 for matter domination, 
$1/3$ for radiation domination), and $e\equiv\chi f_{\chi\chi}/f_\chi$.  
To go further we must adopt a specific form for the function $f$.  Taking 
$f(\chi)=A\chi^n$, we have $e=n-1$ and 
\bea 
\rho_\phi&\sim& a^{-3n(1-w_b)/(2n-1)} \\ 
w_\phi&=&\frac{1-n(1+w_b)}{2n-1} \ . 
\eea 
While in the linear model ($n=1$), the dark energy traces the matter 
during matter domination, this is not so in the nonlinear model.  The case 
$n=0$ is a cosmological constant.  Note that the dark energy is phantom 
(and has a ghost, we will later find) for 
$0<n<1/2$.  This means that to avoid violation of early radiation/matter 
domination the field would have to be highly fine tuned, more so than a 
cosmological 
constant.  For $n\approx 1/2$, the evolution diverges ($e=-1/2)$ and 
matter/radiation domination is violated.  Note that this rules out functions 
that act like $n=1/2$ power laws at early times, such as a DBI type 
$f=\sqrt{1+\chi}-1$.  Therefore we concentrate on $n>1/2$.

\subsection{Late Time Evolution to de Sitter State} \label{sec:late} 

During its evolution, the model leads to cosmic acceleration near the 
present and an asymptotic de Sitter state.  Interestingly, this can arise 
in two ways.  For $\bar{H}'=0$ and also $x' = 0$ as a fixed point one needs 
$\gamma=0$ and either $\omega = 0$ or $\alpha = 0$.  Combining the expression 
for $\gamma$ with Eq.~($\ref{eq:op1}$) leads to the condition 
\begin{equation} 
\label{eq:56} \bar H^2 x^2\,[c_1+f_{\chi}\,(c_{2}+6c_G\bar{H}^{2})]=0 \ . 
\end{equation} 
This guarantees that $\omega = 0$ also.  In the $c_1=0$ case, this is the 
same de Sitter point $\bar H^2_{\rm dS}=-c_2/(6c_G)$ as in \cite{applin} 
and exists irrespective of the 
functional form of $f(\chi)$ (as long as $f_\chi\ne0$ at the de Sitter point). 
Note that $c_{2}$ and $c_{\rm G}$ must have opposite signs for this de Sitter 
point to be present.  In the $c_2=0$ case, there is a new de Sitter point 
$\bar H^2_{\rm dS}=-c_1/(6f_\chi c_G)$. 

Since for the $c_1=0$ de Sitter point we have $\chi\to0$, we should 
choose a function $f$ such that $f(0)=0$ otherwise we are putting in a 
cosmological constant.  But then no solution for this de Sitter point exists 
for the nonlinear power law model that simultaneously satisfies 
$\bar H\to\,{\rm const}$ and $\rho_\phi\to\,{\rm const}\ne0$.  Recall 
that $f_\chi\sim\chi^{n-1}$.  

However, there is yet another de Sitter solution that we can construct for 
our nonlinear model; this arises because of the evolution of $x$ 
such that asymptotically $x'\ne0$.  This solution still has 
$\bar H^2_{\rm dS}=-c_2/(6c_G)$ but $x\sim a^{3(n-1)/(3n-1)}$.  
Thus as $\bar H\to\,{\rm const}$, $x$ decays to 0 for $n<1$ 
while $x$ diverges for $n>1$. 

\begin{figure}[htbp!] 
\includegraphics[angle=-90,width=0.8\textwidth]{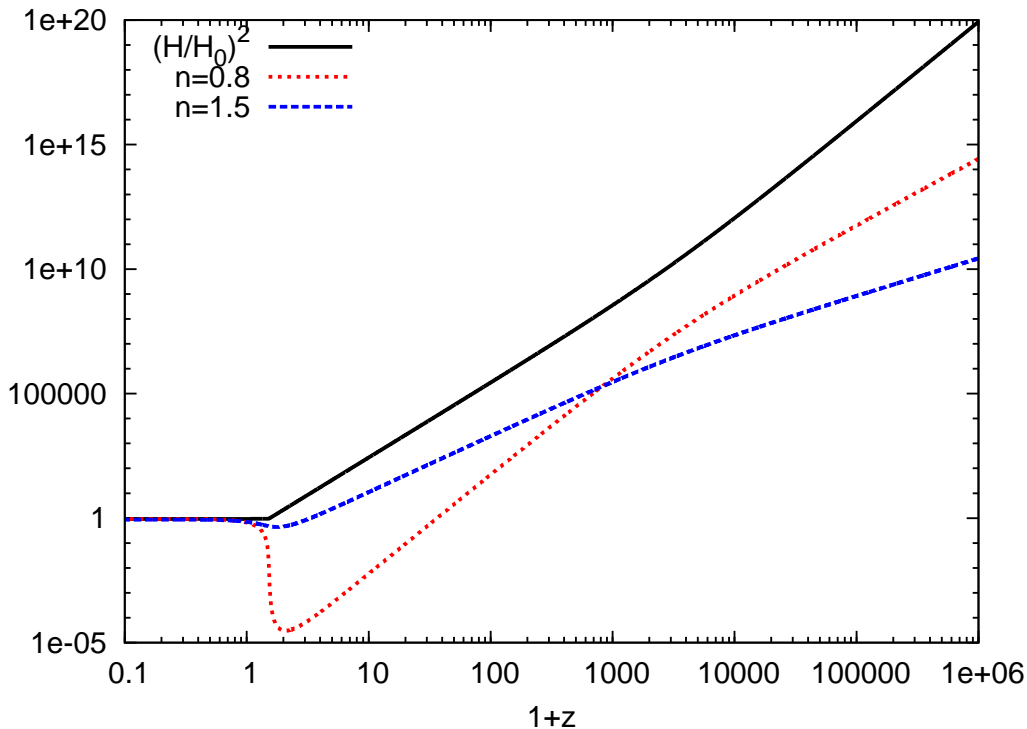} 
\caption{The evolution of the effective dark energy density $\rho_\phi$ 
in the $c_1=0$ case is plotted for power law functions $f\sim\chi^n$ with 
$n=0.8$ (dotted red 
curve) and $n=1.5$ (dashed blue).  The expansion history $(H^{2}/H_{0})^{2}$ 
(solid black) is also shown (for $n=0.8$ though the $n=1.5$ case is nearly 
identical on this scale).  Note that for 
$n<1$, typically $\rho_\phi$ must be set to large values initially so that 
it does not decay to too small values at late times; conversely, for $n>1$ 
$\rho_\phi$ grows relative to matter and radiation and must be set to low 
values initially. 
}
\label{fig:bgdc2} 
\end{figure}

In Fig.~\ref{fig:bgdc2} we exhibit $\bar{H}^{2}$ and $\rho_{\phi}$ for the 
$c_1=0$ cases with $n=1.5$ and $n=0.8$.  As noted, the $n>1$ case grows 
quickly relative to the background components and so must start with a small 
$(\rho_{\phi}/\rho_{m})_i$ to preserve later matter domination.  At 
$a=10^{-6}$, say, this ratio must be less than $10^{-6.6}$, but this is 
still not as severe as the cosmological constant fine tuning which requires 
$10^{-16}$.  The $n=0.8$ case can actually dominate over matter at 
$a=10^{-6}$, but has rather drastic evolution at $z\approx1$ as it suddenly 
turns toward the de Sitter attractor.

\begin{figure}[htbp!]
\includegraphics[angle=-90,width=0.48\textwidth]{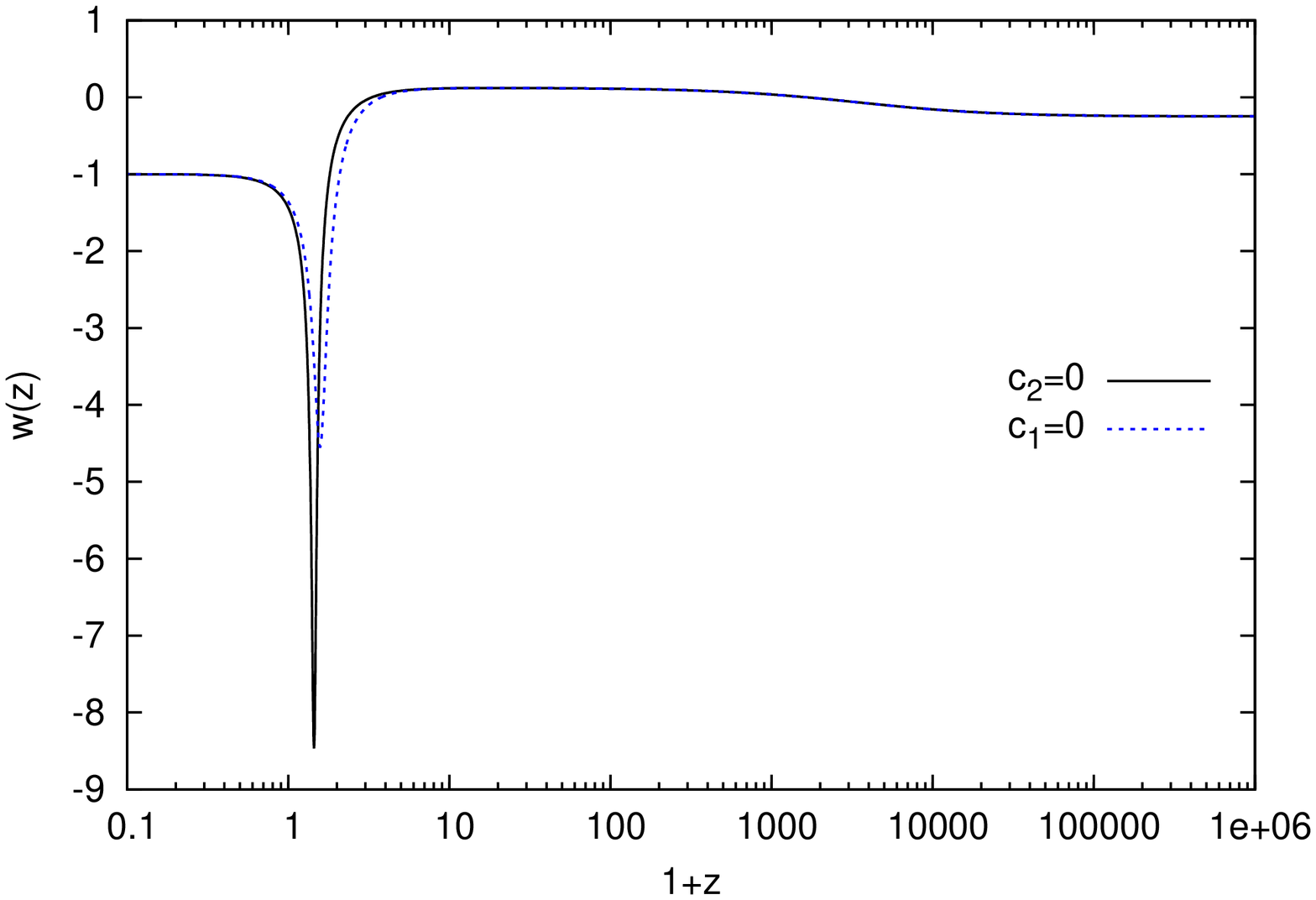}
\includegraphics[angle=-90,width=0.48\textwidth]{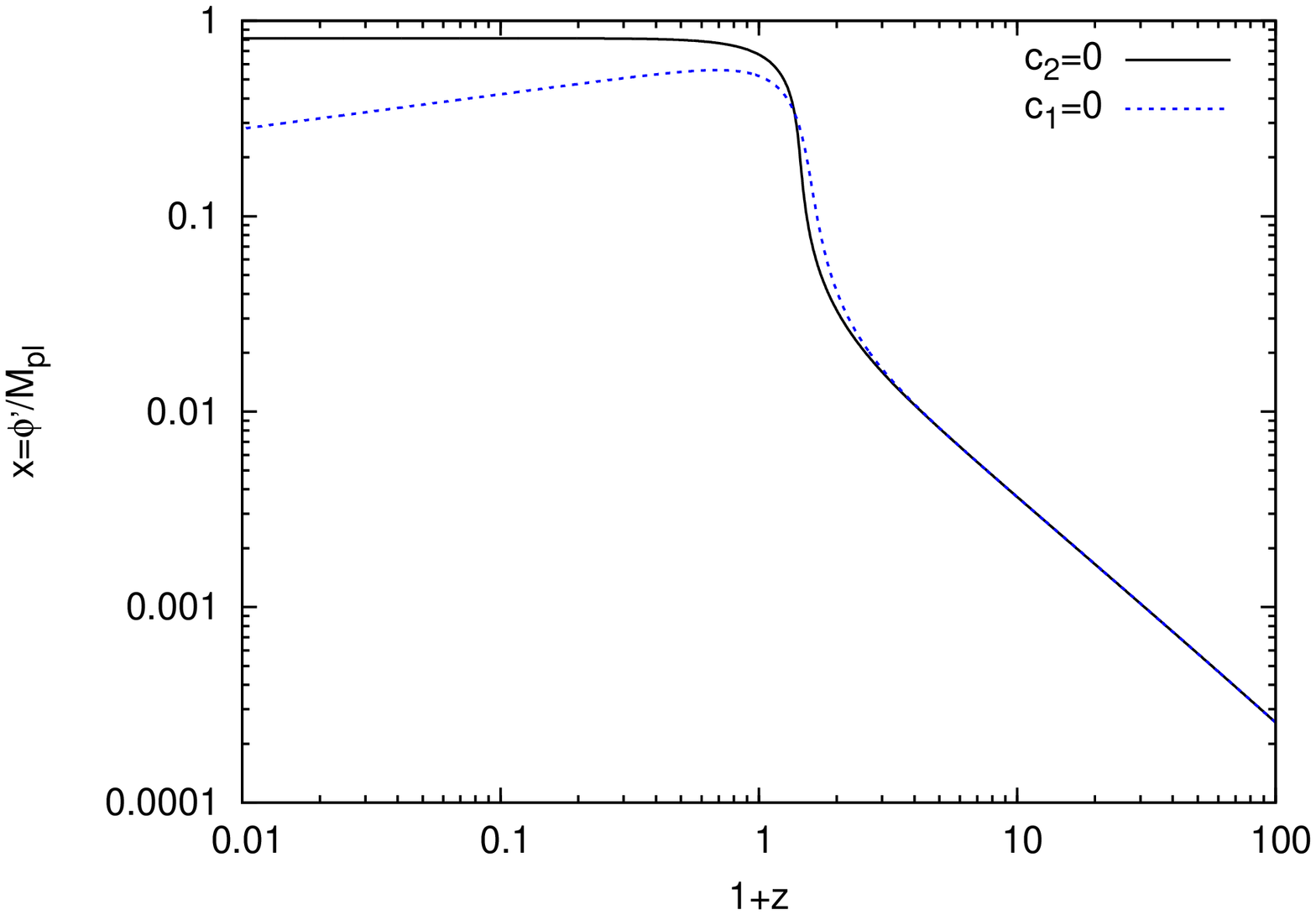}
\caption{Comparison of the cases $c_{2}=0$ (solid black) and $c_{1}=0$ 
(dotted blue), both with $n=0.9$ and 
$[\rho_\phi/\rho_m](a=10^{-6})=10^{-3}$, is plotted for the 
equation of state parameter (left panel) and the field evolution 
$x = \phi' /M_{\rm pl}$ (right panel) of the effective dark energy.  
The $c_2=0$ 
case has a more extreme phantom feature in $w(z)$ near the present, but 
both cases have the same early time tracking and late time de Sitter 
attractor (although for $c_1=0$ one has $x'\ne0$). 
} 
\label{fig:wc2c1}
\end{figure}

For the $c_2=0$ case, as mentioned we expect at early times no significant 
change to the dynamics since again the $c_{\rm G}$ term will dominate over 
$c_1$.  At late times, since $\rho_\phi$ contains terms with different 
powers of $x$ there is no extra de Sitter solution (hence 
$\rho_\phi=\,{\rm const}$) with varying $x$.  Thus the only de Sitter 
solution is $c_1+6f_\chi c_G\bar H^2=0$.  Note that $\chi$ freezes at a 
finite value and so $f\to\,{\rm const}$.  The energy density evolution 
looks quite similar to the $c_1=0$ case and so instead we show the 
evolution of the dark energy equation of state for the two cases in 
the first panel of Fig.~\ref{fig:wc2c1}.  The spike in $w_\phi(z)$ can be 
ameliorated by raising the initial field energy density, but as discussed 
above this would impinge on matter domination.

Although both cases reach de Sitter attractors asymptotically, the 
manner in which they achieve this differs.  For the nonlinearity 
only applying to the Einstein tensor coupled kinetic term ($c_2=0$ case), 
the solution is the double fixed point $\bar H'=0=x'$.  This holds as 
well for the full nonlinearity (applied to both kinetic terms, i.e.\ the 
$c_1=0$ case) when $n>1$ (but this is not the stable attractor).  
In addition the full nonlinearity case also 
has a de Sitter solution with $\bar H'=0$ but $x'\ne0$, i.e.\ 
$\ddot\phi\ne0$.  The field will decelerate, $\ddot\phi<0$ (accelerate, 
$\ddot\phi>0$) for $n<1$ ($n>1$).  The difference between the field 
evolutions for the two 
cases is shown in the second panel of Fig.~\ref{fig:wc2c1}, for $n=0.9$. 

A summary of the de Sitter attractors is given in Table~\ref{tab:c1c2}, 
including results from Sec.~\ref{sec:perturb} on the ghost and Laplace 
stability conditions of the perturbations.

\begin{table}[!htb]
\begin{tabular}{c|c|ccccc} 
$\quad (c_1,c_2)\quad$&$\quad$ $n$&$H^2_{\rm dS}$&$x_{\rm dS}$&$\chi_{\rm dS}$& 
No-ghost& Laplace\\ 
\hline 
$(0,c_2)$&$1/2<n<1$&$-c_2/(6c_G)$&0&0&$\checkmark$&$\sim\checkmark$\\ 
$(0,c_2)$&$n>1$&$-c_2/(6c_G)$&$\infty$&$\infty$&$\times$&$\times$\\ 
$\quad(c_1,0)\quad$&$\quad 1/2<n<1\quad$&$\quad -c_1/(6c_G f_\chi)\quad$& 
$\quad$const$\quad$&$\quad$const$\quad$&$\checkmark$&$\sim\checkmark$\\ 
$(c_1,0)$&$n>1$&$-c_1/(6c_G f_\chi)$&const&const&$\times$&$\times$ 
\end{tabular}
\caption{Summary of de Sitter attractors is given for the two cases of 
the model, with a nonlinear function $f\sim\chi^n$.  The two cases have 
different approaches to de Sitter, that merge in the common linear 
limit $n=1$.} 
 \label{tab:c1c2}
\end{table}

\section{Self Tuning} \label{sec:tune} 

The ability of the theory to reach a de Sitter asymptotic state without 
a cosmological constant is interesting, as is the overall expansion 
behavior of such a cosmological model, but 
more significant is the ability of Fab 5 Freddy to self tune, in the manner 
of John or Paul in the Fab Four \cite{fab4}.  This allows the scalar field 
$\phi$ -- even without a potential -- to cancel an 
existing (large) cosmological constant.  This even holds if the cosmological 
constant readjusts as it passes through phase transitions.  Here we present 
a simplified analysis showing these key properties while neglecting matter 
or radiation components. 

The dynamical equations are identical to Eqs.~(\ref{eq:67})-(\ref{eq:chi}) 
except for the replacement of $\Omega_{r0}\,a^{-4}$ by $-3\Omega_{\Lambda}$ 
in the $\gamma$ term, coming from (three times) the background pressure. 
(Note $\Omega_{\Lambda}\ne0.7$, the observed cosmological constant, but 
is instead the early universe, bare cosmological constant.)  
Two de Sitter points can be found, both of which are attractors.  The first 
arises from the explicit cosmological constant, with $\bar{H}_{1}^{2} = 
8\pi G\rho_{\Lambda}/(3H_{0}^{2}) = \Omega_{\Lambda}$ and the second is 
the self tuning solution with $\bar{H}_{2}^{2} = -c_{2}/(6c_{\rm G})$, as 
we found in the absence of a cosmological constant. 
Note $\bar{H}_{1} \gg \bar{H}_{2}$.

For the first solution, the scalar field contribution dies away as 
$\rho_\phi \sim x^{2n} \sim a^{-6n/(2n-1)}$, so the pure cosmological 
constant is a fixed point of the dynamics.  (Of course matter and radiation 
would also redshift away.)  For the second solution the scalar field 
dynamically adjusts such that $\rho_{\phi} \to -\rho_{\Lambda}$.  Note 
that unlike in the earlier sections $\rho_\phi<0$.  However the same 
approach to a de Sitter state occurs, with $\rho_\phi$ dynamically 
canceling $\rho_\Lambda$ and retaining a small positive residual energy 
density, evolving with $x \sim a^{-3(n-1)/(3n-1)}$ on approach to 
$\bar{H}^{2}_{2} \to -c_{2}/(6c_{\rm G})$. 

\begin{figure}[htbp!]
\includegraphics[angle=-90,width=0.49\textwidth]{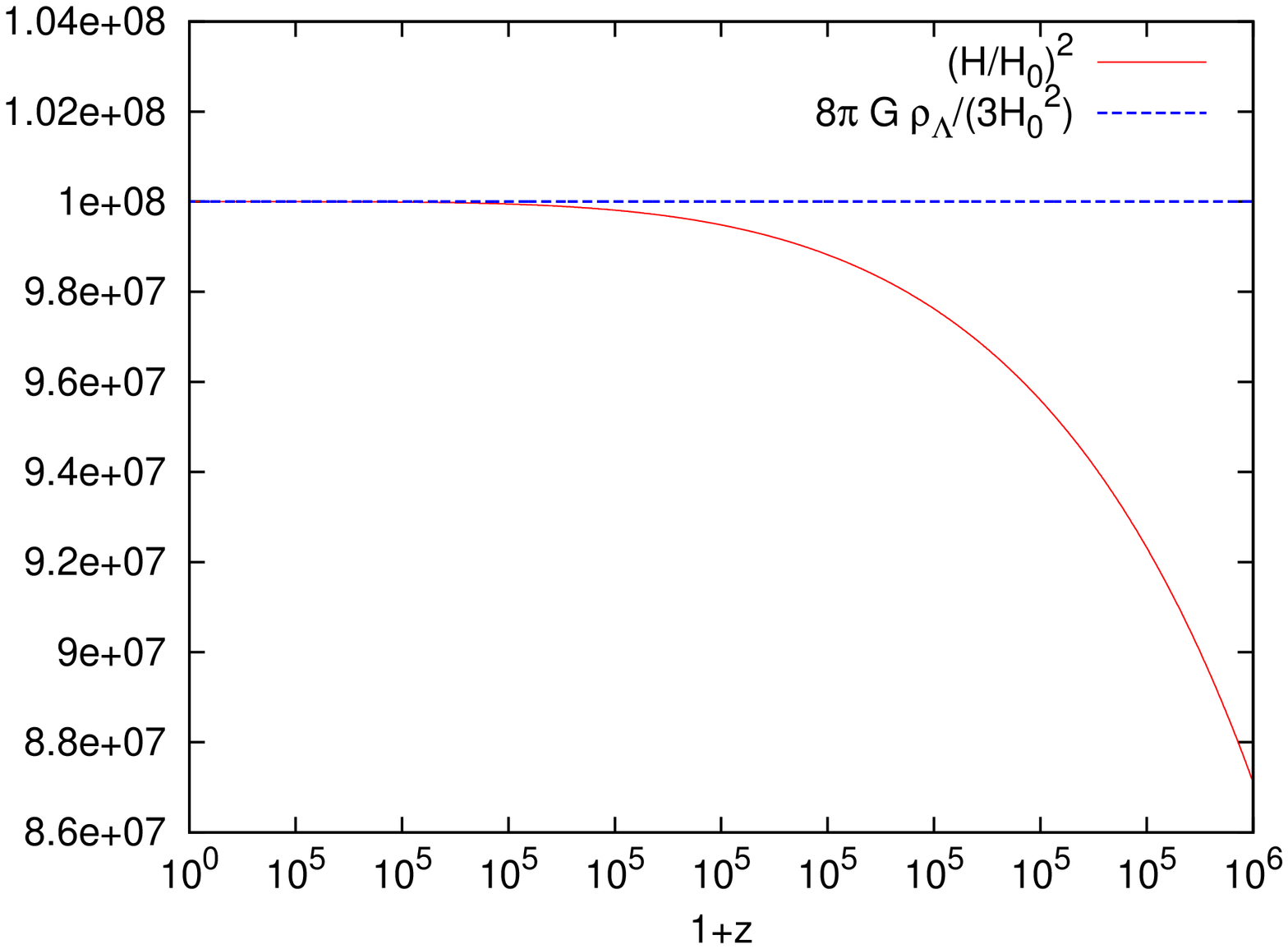}
\includegraphics[angle=-90,width=0.49\textwidth]{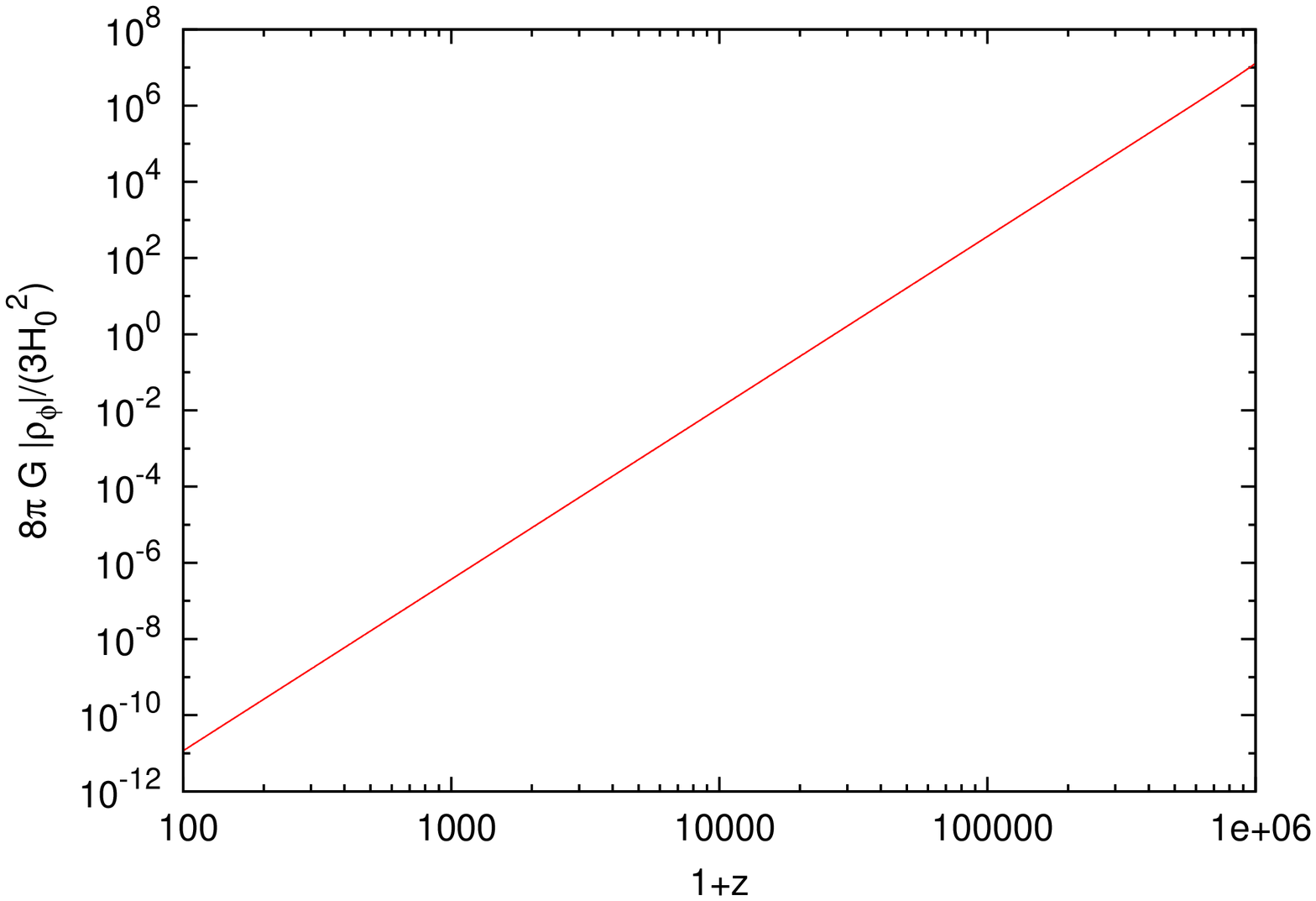}
\caption{  [Left panel] $\bar{H}^{2}$ evolves toward its standard 
cosmological constant attractor 
$\bar{H}^{2}_1 = 8\pi G \rho_{\Lambda} /(3H_{0}^{2})$ 
for high density initial conditions, here $\bar H_i = 10^{4}\sqrt{-c_{2}/ 
(6c_{\rm G})}$.  [Right panel] Meanwhile the scalar field energy density 
decays away.}
\label{fig:st1}
\end{figure}

\begin{figure}[htbp!]
\includegraphics[angle=-90,width=0.49\textwidth]{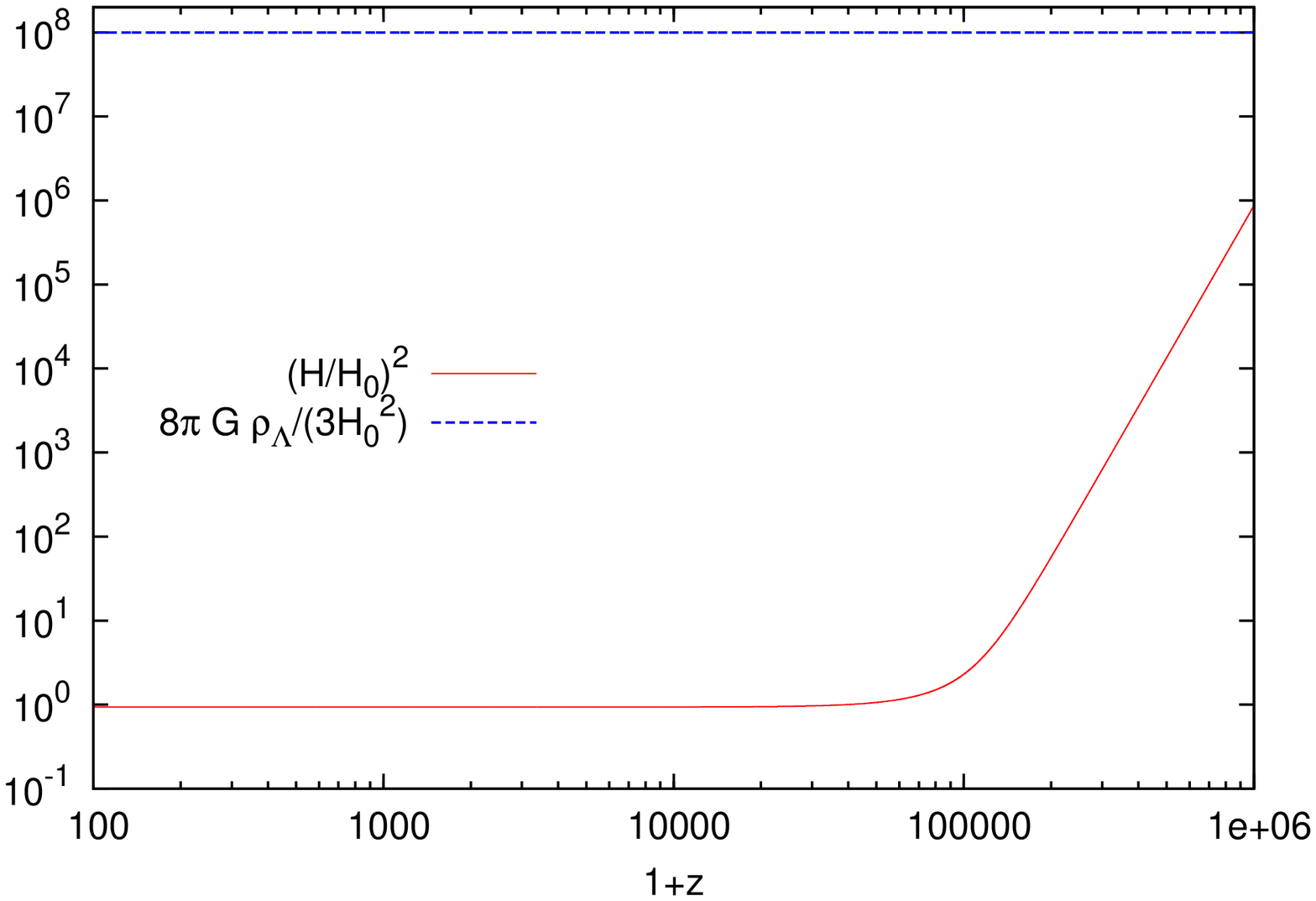}
\includegraphics[angle=-90,width=0.49\textwidth]{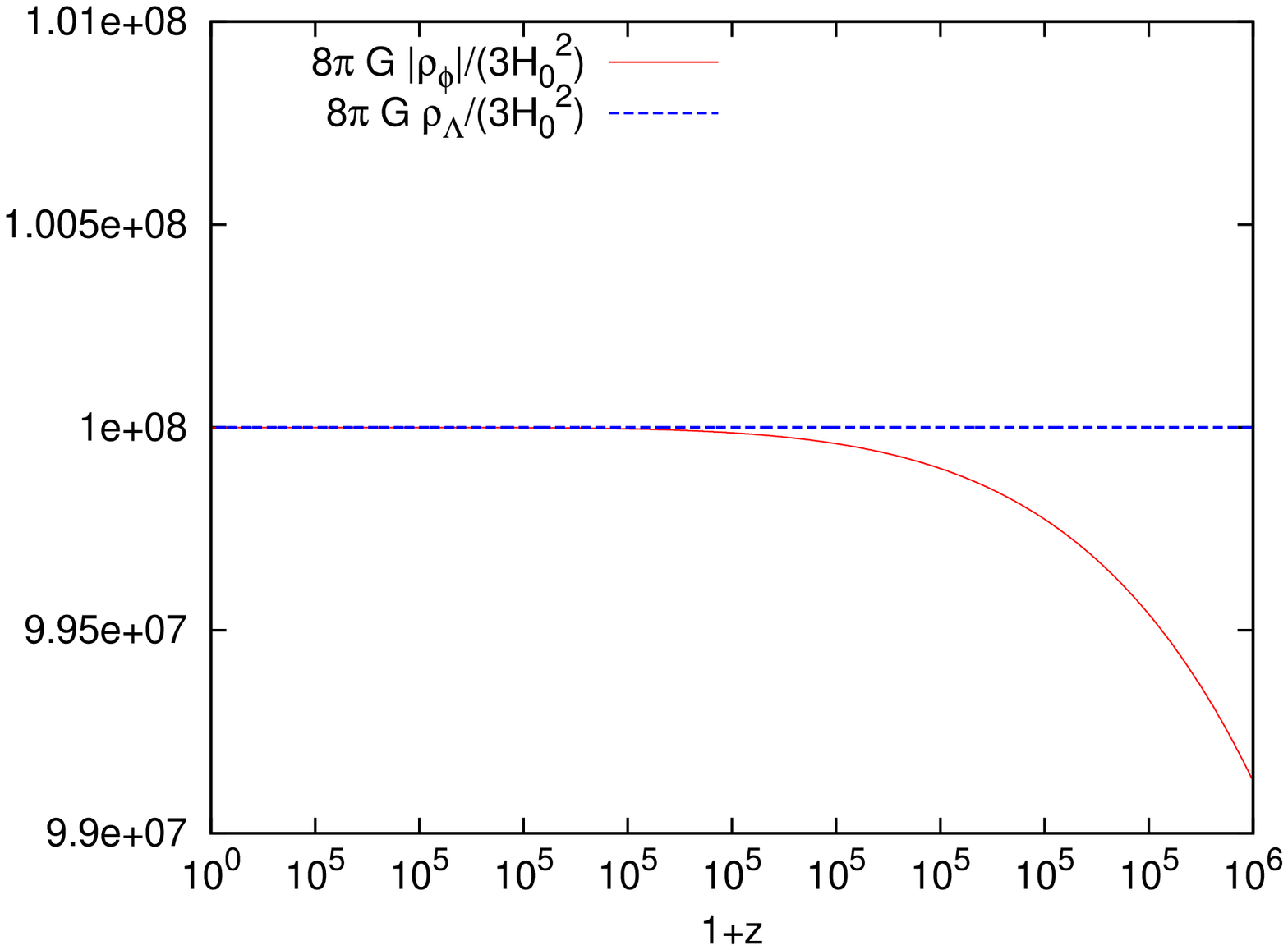}
\caption{As Fig.~\ref{fig:st1}, but with initial conditions 
$\bar H_i = 10^{3}\sqrt{-c_{2}/(6c_{\rm G})}$.  [Left panel] Now we 
observe the dynamics leads to the second attractor $\bar{H}^{2} \to 
-c_{2}/(6c_{\rm G})$, despite the presence of a large cosmological constant. 
[Right panel] The scalar field energy density $\rho_\phi$ self tunes to 
cancel the $\rho_{\Lambda}$ contribution in the field equations. } 
\label{fig:st2}
\end{figure}

These analytic behaviors are verified numerically in Figs.~\ref{fig:st1} 
and \ref{fig:st2}.  We include a cosmological constant 
$\Omega_{\Lambda} = 10^{8}$ throughout the numerical calculation, and 
consider $f(\chi) = -\chi^{n}$ (adopting $n=1.5$, $c_{\rm G} = 1$, 
$c_{2}=-5.6$, $c_{1}=0$). We begin $\bar{H}$ away from both asymptotic 
solutions $\bar{H} = \bar{H}_{1,2}$, illustrating the behaviors for 
different initial conditions in Figs.~\ref{fig:st1} and \ref{fig:st2}. 

In the left panel of Fig.~\ref{fig:st1} we observe the approach to the 
standard cosmological constant attractor with $\bar{H}^{2} \to 8\pi G 
\rho_{\Lambda}/(3H_0^2)$, with the right panel showing the vanishing 
$\rho_{\phi} \to 0$. 

However, below a certain critical initial condition $\bar H_i$ (depending 
on the other parameters), we observe entirely different dynamical behavior. 
Now, $\bar{H}$ approaches the second asymptotic point $\bar{H}_2^2 = -c_{2}/ 
(6c_{\rm G})$.  This occurs despite the large cosmological constant 
present in the model. We find that the absolute value of the $\phi$ field 
energy density approaches the $\rho_{\phi} \sim -\rho_{\Lambda}$ solution, 
canceling the vacuum energy in the field equations. Hence the model 
exhibits self tuning, for some range of initial conditions 
(e.g.\ $\bar H^2(a=10^{-6})\lesssim 10^{6} \bar H_2^2$ for the parameters 
adopted in the figure).

To see how self tuning occurs, we must examine the equations of motion (here 
taking $c_{1}=0$) 
\begin{eqnarray} 
3\Mpl^{2}\,{H}^{2}&=& 
\frac{1}{2}c_{2}f_\chi \dot{\phi}^{2}+f_\chi\,\chi-f+ 
9{c_{G} \over M^{2}} f_\chi {H}^{2}\dot{\phi}^{2}\ , \\
\left(c_{2}+6 {c_{G}\over M^{2}}{H}^{2}\right)f_\chi \,\ddot{\phi}&=& 
-12{c_{G} \over M^{2}} f_\chi\,\dot{\phi} H\dot{H}-
(\dot{f_\chi}+3 Hf_\chi)\left(c_{2}+6 {c_{G}\over M^{2}}{H}^{2}\right)\dot{\phi} \ ,\\ 
\chi&=& \left(\frac{1}{2}c_{2}+3{c_{G} \over M^{2}}H^{2}\right)\dot{\phi}^{2}\ . 
\end{eqnarray} 
On-shell (that is, at the asymptotic de Sitter state), we have $\bar{H}^{2} = -c_{2}/(6c_{\rm G})$ and $\dot{H} = 0$, and hence the scalar field equation is 
trivially satisfied, carrying no 
information regarding the evolution of $\phi$. However, 
the scalar field equation contains an explicit $\ddot{a}$ dependence, and 
the Hamiltonian density ${\cal H}$ in the Friedmann equation retains 
$\dot{\phi}$ dependence on shell, 
both of which are conditions given in \cite{fab4long} for self tuning to 
occur. 
On approach to the de Sitter point, the scalar field continues to 
evolve while $\rho_{\phi}$ and $\bar{H}$ approach constant values.

If we choose initial conditions for $\bar{H}_i$ such that it is initially far 
from the attractor $\bar{H}_2^{2} = -c_{2}/(6c_{\rm G})$, then one can use 
Eqs.~(\ref{eq:67}) and (\ref{eq:68}) to calculate how the model approaches the de Sitter state. For the power law models $f(\chi) \sim \chi^{n}$, we find 
that the dynamical behaviour of $\bar{H}$ is independent of $n$, and the 
evolution toward de Sitter has $\bar{H}-\bar{H}_{\rm dS} \sim a^{-3}$, 
$x \sim a^{6}$. Ultimately the evolution of $\bar{H}$ will depend on the presence of matter and radiation (which we have neglected here), and also the functional form of $f(\chi)$. 
Whether a specific self tuning model can be constructed that gives rise to 
a viable cosmological evolution will be the subject of future work (see 
\cite{fab4new} for the Fab Four case).

\begin{figure}[htbp!]
\includegraphics[angle=-90,width=0.49\textwidth]{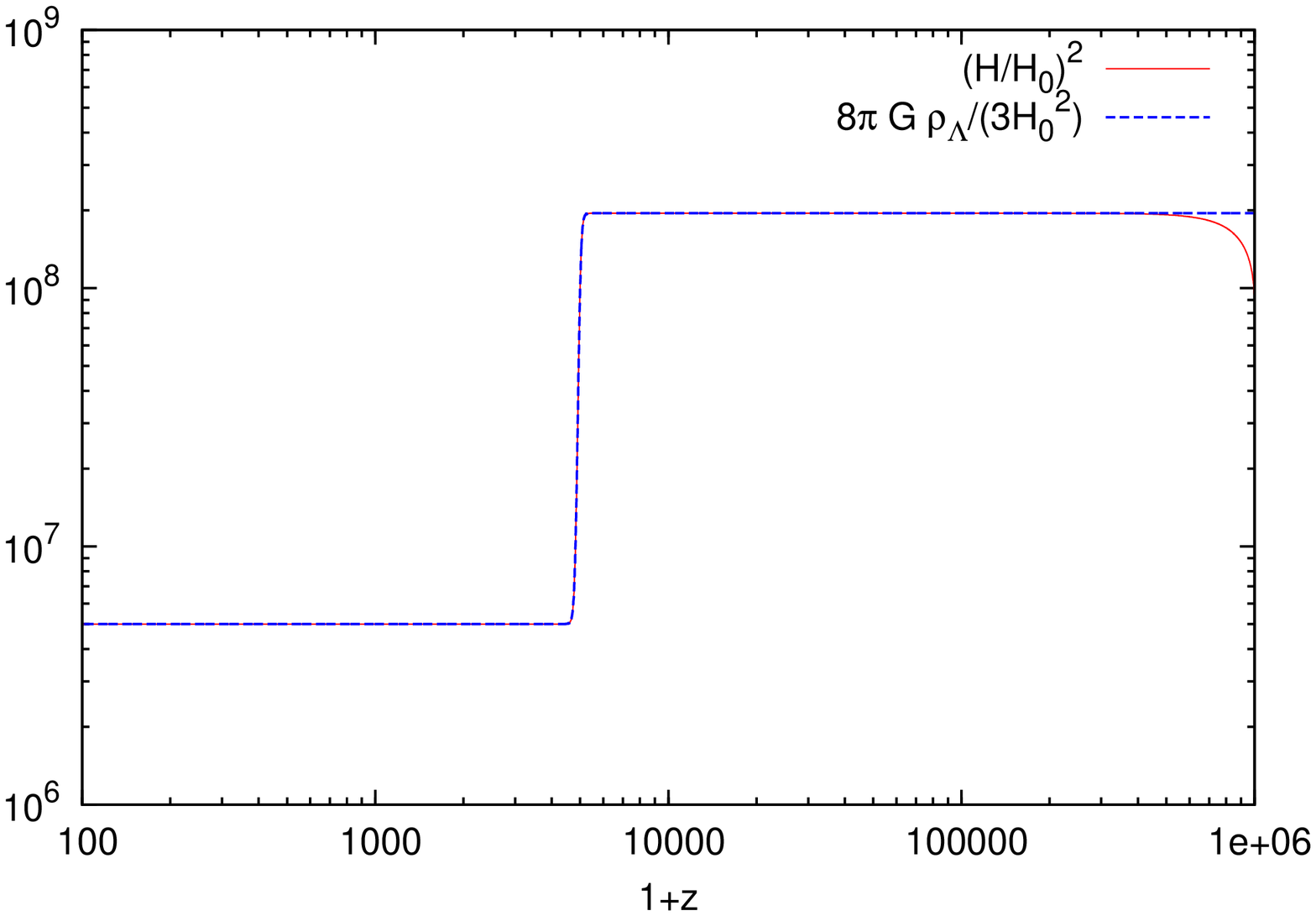}
\includegraphics[angle=-90,width=0.49\textwidth]{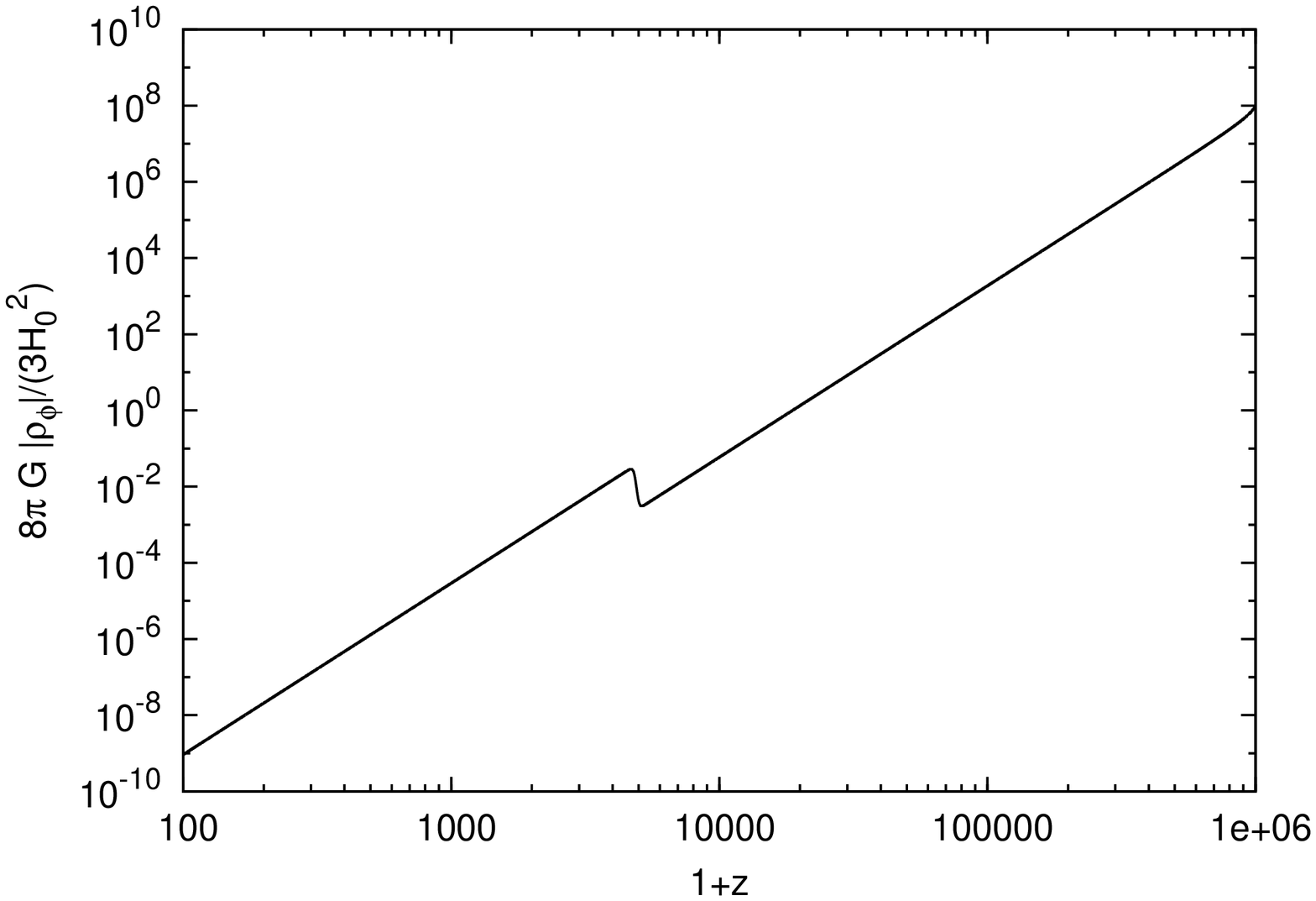}
\caption{As Fig.~\ref{fig:st1} for $\bar H_i = 10^{4}\sqrt{-c_{2}/(6c_{\rm G})}$, 
but with the large vacuum energy undergoing a phase transition.  The 
standard attractor $\bar{H}^{2} \to 8\pi G \rho_{\Lambda}/(3H_{0}^{2})$ 
applies, and the $\phi$ field energy density asymptotically decays, subject 
to a mild jump at the phase transition.}
\label{fig:st3}
\end{figure}

\begin{figure}[htbp!]
\includegraphics[angle=-90,width=0.49\textwidth]{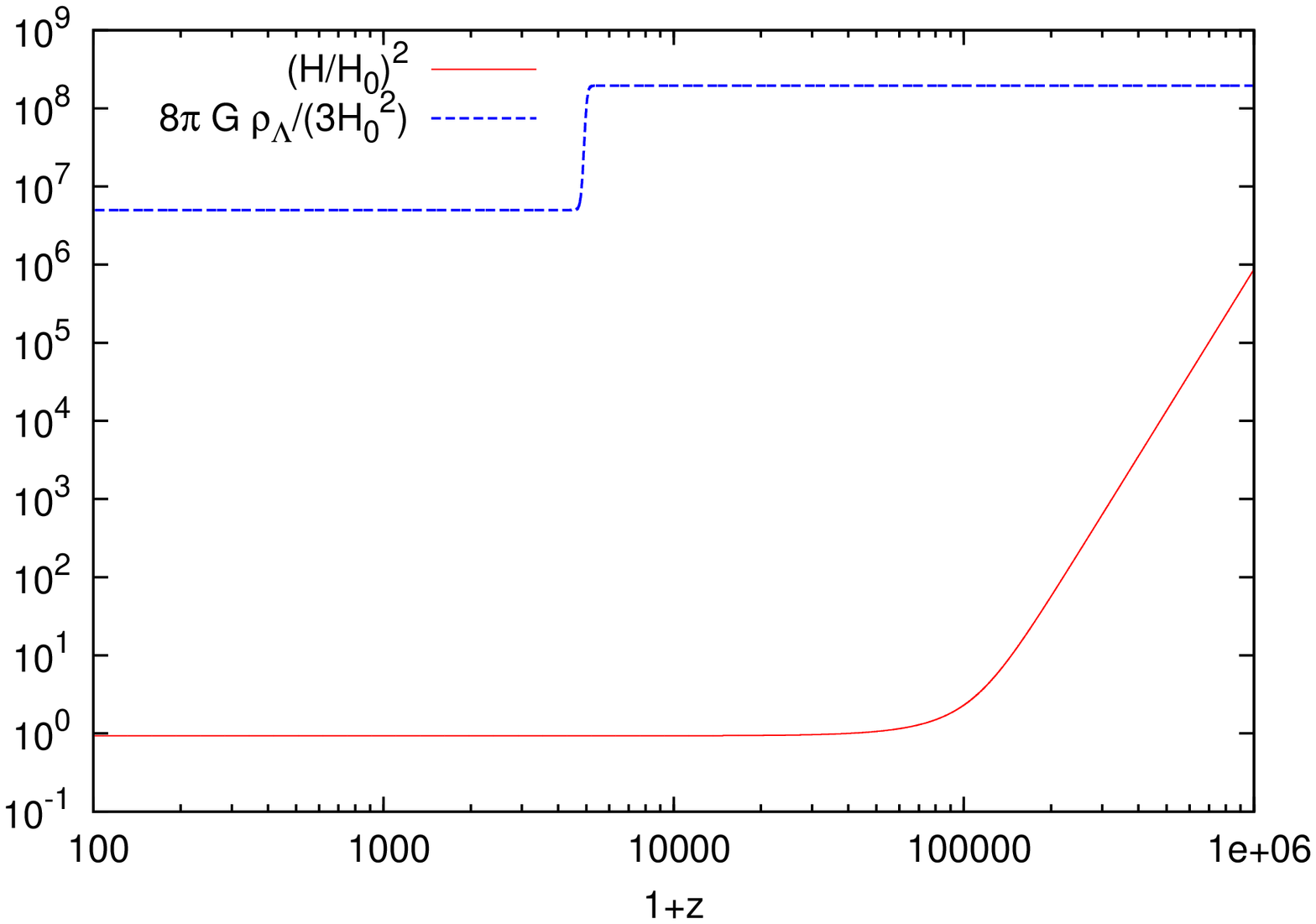}
\includegraphics[angle=-90,width=0.49\textwidth]{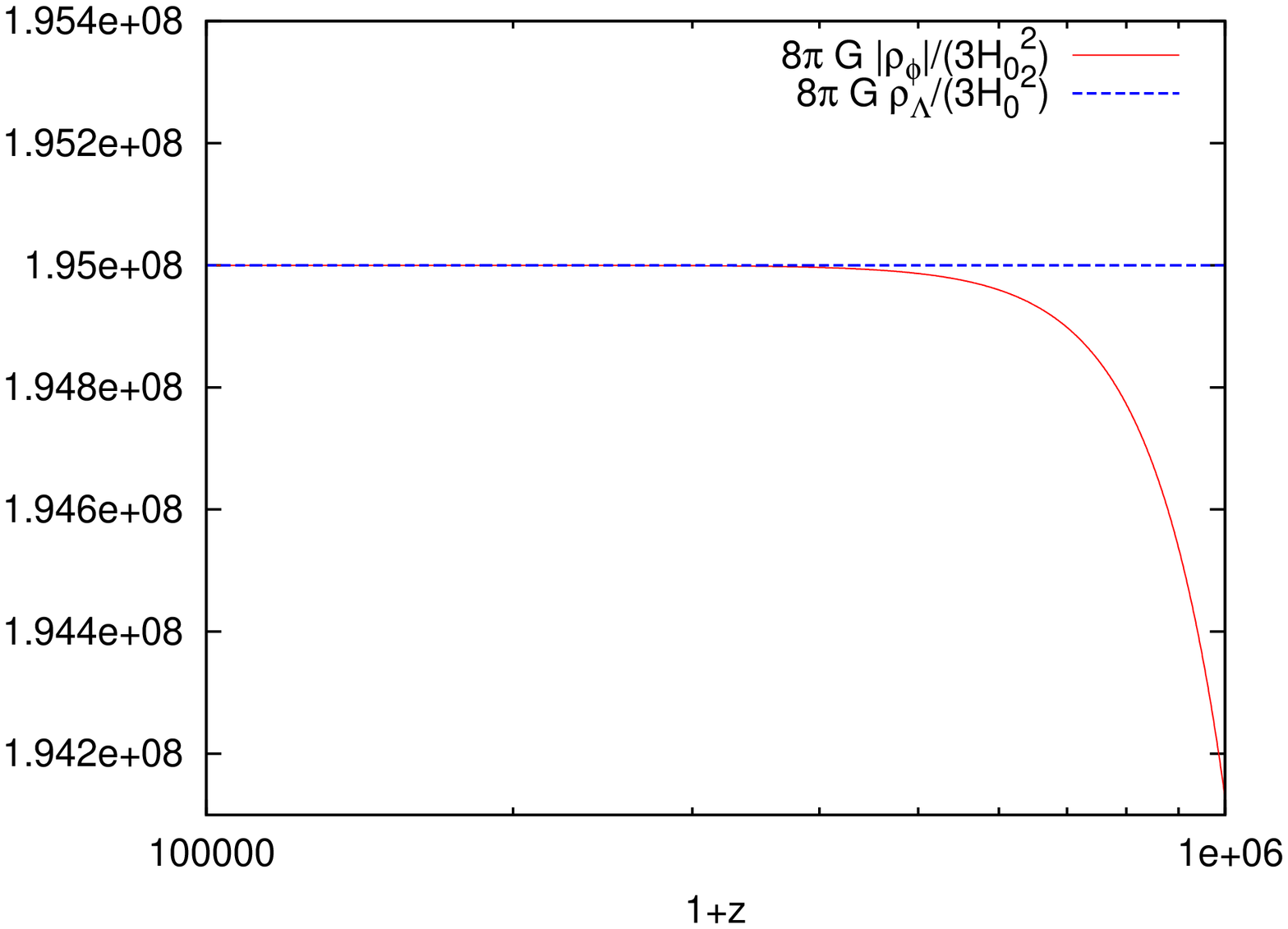}
\caption{As Fig.~\ref{fig:st3} with a large vacuum energy undergoing a 
phase transition, but with $\bar H_i = 10^{3}\sqrt{-c_{2}/(6c_{\rm G})}$. 
Self tuning remains effective despite the transition and the second attractor 
$\bar{H}^{2} \to -c_{2}/(6c_{\rm G})$ is approached.  The $\phi$ field 
dynamically adjusts energy density to cancel the vacuum energy, and the 
transition does not modify the Hubble parameter due to it already being on 
the attractor. } 
\label{fig:st4}
\end{figure}

Going further, we can verify that the self tuning also self adjusts if the 
vacuum energy undergoes a phase transition at some redshift.  We numerically 
model such an energy density with a tanh function, and choose the pressure 
to solve the continuity equation $P = -{\rho' \over 3} - \rho$.  
The evolution of the quantities 
$\bar{H}^{2}$, $\rho_{\phi}$, and $\Omega_{\Lambda}$ are shown in 
Figs.~\ref{fig:st3} and \ref{fig:st4}, demonstrating that the two de Sitter 
solutions still hold and the self tuning mechanism remains effective. 
The explicit cosmological constant can be made effectively invisible in 
our model.

\section{Linear Perturbations} \label{sec:perturb} 

Linear perturbations of the equations of motion are important for 
calculating the growth of structure and assessing the ghost-free and 
stability conditions of the theory.   For subhorizon perturbations one 
adopts the quasistatic approximation.  We begin by using Eq.~(\ref{eq:actio2}) 
to write the equations of motion for the perturbation in the Newtonian gauge 
($\beta=0$ in Appendix~\ref{sec:apxpert}) as follows: 
\begin{equation} 
2M_{\rm pl}^{2} \nabla^{2} \Phi =  \rho_{\rm m}\delta_{\rm m} + 9 {c_{\rm G} \over M^{2}} f_{\chi\chi} H^{2} \dot{\phi}^{2}\delta \chi + {c_{\rm G}\over M^{2}} f_{\chi} \left( 2 \dot{\phi}^{2} \nabla^{2}\Phi - 4 H \dot{\phi} \nabla^{2} \delta \phi \right) +  f_{\chi\chi}\chi \delta \chi + {c_{2} \over 2} f_{\chi\chi}\dot{\phi}^{2} \delta \chi 
\end{equation} 
\begin{equation} 
\delta \chi = 2 {c_{\rm G} \over M^{2}} \dot{\phi}^{2} \nabla^{2} \Phi 
\end{equation}
\begin{eqnarray} 
\label{eq:pipert}  & &  c_{1} \nabla^{2}\delta \phi + c_{2} f_{\chi} \nabla^{2}\delta \phi - c_{2} \dot{\phi} (f_{\chi\chi}\delta \chi)\dot{} + 2{c_{\rm G} \over M^{2}} f_{\chi} (2\dot{H} + 3H^{2}) \nabla^{2} \delta \phi - 6 {c_{\rm G} \over M^{2}} H^{2} \dot{\phi}(f_{\chi\chi}\delta \chi)\dot{} - c_{2} f_{\chi\chi} (\ddot{\phi} + 3H\dot{\phi})\delta\chi 
\\  \nonumber
& & - 6 {c_{\rm G} \over M^{2}} f_{\chi\chi} H^{2} \ddot{\phi} \delta \chi - 6 {c_{\rm G} \over M^{2}} f_{\chi\chi}H(2\dot{H} + 3H^{2})\dot{\phi} \delta \chi - 4 {c_{\rm G}\over M^{2}} f_{\chi} \ddot{\phi} \nabla^{2} \Phi - 4 {c_{\rm G} \over M^{2}} \dot{f}_{\chi} \dot{\phi} \nabla^{2}\Phi + 4 {c_{\rm G} \over M^{2}} f_{\chi} H \dot{\phi} (\nabla^{2} \psi -\nabla^{2}\Phi) = 0 
\end{eqnarray}  
\begin{eqnarray} 
\nonumber & & \partial_{i}\partial_{j} \Phi - \partial_{i}\partial_{j}\psi + g_{ij} (\nabla^{2} \psi - \nabla^{2}\Phi) =  {c_{\rm G} \over M^{2}} f_{\chi} \left\{ 2 (\ddot{\phi} + H\dot{\phi}) (g_{ij}\nabla^{2} \delta \phi - \partial_{i}\partial_{j}\delta \phi )  + \dot{\phi}^{2} \left[ g_{ij} \left( \nabla^{2} \Phi + \nabla^{2} \psi\right) - (\partial_{i}\partial_{j}\Phi + \partial_{i}\partial_{j}\psi) \right]\right\} \\ 
\nonumber & & + g_{ij} \left( {c_{2} \over 2} f_{\chi\chi} \dot{\phi}^{2} \delta \chi - {c_{\rm G} \over M^{2}} \left[ 2 H \dot{\phi}^{2} \left( f_{\chi\chi}\delta \chi\right)\dot{} 
+ f_{\chi\chi}\left( 4H\dot{\phi}\ddot{\phi} + 2 \dot{H} \dot{\phi}^{2} + 3 H^{2} \dot{\phi}^{2} \right) \delta \chi \right] - f_{\chi\chi}\chi\delta\chi \right) \\ 
& & + {c_{\rm G} \over M^{2}}  \dot{\phi}^{2} f_{\chi\chi} \left( \partial_{i}\partial_{j} \delta \chi - g_{ij} \nabla^{2} \delta \chi \right)  + 2{c_{\rm G} \over M^{2}} \dot{f}_{\chi}\dot{\phi} (g_{ij} \nabla^{2} \delta \phi - \partial_{i}\partial_{j} \delta \phi ) \ . 
\end{eqnarray} 

There are some important differences between this case and the linear 
case where $f_{\chi\chi}=0$, i.e.\ no nonlinear mixing.  Here, in the 
$(i,j)$ Einstein and $\phi$ field equations, terms appear of the form 
$k^{4} \Phi f_{\chi\chi}$ and $k^{2} \dot{\Phi} f_{\chi\chi}$, arising 
from $\delta\chi$.  These will lead to scale dependence in the gravitational 
coupling strength $G_{\rm eff}$ derived below.  Recall that the standard 
Galileon case does not have scale dependent coupling on cosmic scales well 
above the Vainshtein scale (see $\geff$ from \cite{applin}).

\subsection{Evolution of Gravity} \label{sec:geff} 

To investigate the modified Poisson equations defining the coupling of 
matter to the metric potentials, we can use the $(i,j=i)$ perturbed 
Einstein equation to remove $(f_{\chi\chi}\delta\chi)\dot{}$, 
and then substitute for the $(i,j\ne i)$, $\phi$, and $\chi$ equations.  
In the quasistatic limit appropriate for linear growth on subhorizon 
scales the $(0,0)$ perturbed Einstein equation becomes 
\begin{equation} 
{\nabla}^{2} \Phi = {4\pi a^2 G^{(\Phi)}_{\rm eff} \rho_m } \delta_m  \ . 
\end{equation} 
The equivalent modified Poisson equations for the other metric potential 
combinations are 
\begin{eqnarray} 
& & {\nabla}^{2} \psi = {4\pi G_{\rm eff}^{(\psi)} \rho_m} \delta_{\rm m} \\ 
& & {\nabla}^{2} (\Phi+ \psi) = {8\pi G_{\rm eff}^{(\Phi + \psi)} \rho_m } \delta_{\rm m} \ .
\end{eqnarray} 

The gravitational couplings are 
\begin{equation} 
{G_{\rm eff}^{(\Phi)} \over G_{\rm N}} = \frac{\kap{3}\kap{8}+2\kap{2}\kap{9}}{\kap{1}(\kap{3}\kap{8}+2\kap{2}\kap{9}) + \kap{2} (\kap{5}\kap{8}+2\kap{2}\kap{7}-\kap{4}\kap{8}\kap{6}) } \, ,
\label{eq:Geff1}
\end{equation} 
\bea 
{G_{\rm eff}^{(\psi)} \over G_{\rm N}} &=& -\left[ {\kap{9} \over \kappa_{8}} \left({ \kap{1} \bar{G}_{\rm eff}^{(\Phi)} -1 \over \kap{2}}\right) + {\kap{7} \over \kap{8}} \bar{G}_{\rm eff}^{(\Phi)}\right] \\ 
{G_{\rm eff}^{(\Phi+\psi)} \over G_{\rm N}} &=&   \left({\kappa_{8}-\kappa_{7} \over 2\kappa_{8}}\right) \bar{G}_{\rm eff}^{(\Phi)} - {\kap{9} \over 2\kappa_{8}}  \left({ \kap{1} \bar{G}_{\rm eff}^{(\Phi)} -1 \over \kap{2}}\right) \ , \label{eq:Geff3}
\eea 
where $\bar{G}_{\rm eff}^{(\Phi)} = G_{\rm eff}^{(\Phi)}/G_{\rm N}$ and 
\begin{eqnarray} 
& & \kap{1} = 1 - 12c_{\rm G}^{2} \bar{f}_{\chi\chi}\bar{H}^{6} x^{4} - c_{2}c_{\rm G} \bar{f}_{\chi\chi} \bar{H}^{4}x^{4} - c_{\rm G} \bar{f}_{\chi}\bar{H}^{2}x^{2} \\ 
& & \kappa_{2} = -2c_{\rm G} \bar{f}_{\chi} \bar{H}^{2}x \\ 
& & \kappa_{3} = c_{1} + c_{2} \bar{f}_{\chi} + 2 c_{\rm G} \bar{f}_{\chi} \left( 2\bar{H}\bar{H}' + 3\bar{H}^{2} \right) \\ 
& & \kappa_{4} = -c_{2} \bar{H} x - 6 c_{\rm G} \bar{H}^{3}x \\ 
\nonumber & & \kappa_{5} = -4 c_{\rm G} \bar{f}_{\chi} \bar H \left( \bar{H} x' + \bar{H}' x + \bar{H} x \right) - 12 c_{\rm G}^{2} \bar{f}_{\chi\chi} \bar{H}^{5} x^{2} \left( 3\bar{H}x' + 7\bar{H}' x + 3\bar{H}x \right) \\ 
& & \qquad\qquad - 6 c_{2} c_{\rm G} \bar{f}_{\chi\chi} \bar{H}^{3} x^{2} \left( \bar{H} x' + \bar{H}' x + \bar{H}x \right) \\ 
& & \kappa_6 = -c_G f_{\chi\chi} {\bar H}x \left[xk^2+\bar H \left( 4 \bar{H}x'  + 6 \bar{H}' x + 7 \bar{H} x \right) \right] \\ 
& & \kappa_{7} = 1 + c_{\rm G} \bar{f}_{\chi} \bar{H}^{2} x^{2} - 2 c_{\rm G}^{2} \bar{f}_{\chi\chi}\bar{H}^{4}x^{4} k^{2}/a^{2} \\ 
& & \kappa_{8} = -1 + c_{\rm G}\bar{f}_{\chi}\bar{H}^{2}x^{2} \\ 
& & \kappa_{9} = 2c_{\rm G} \bar{f}_{\chi} \bar{H} \left( \bar{H} x' + \bar{H}' x + \bar{H} x \right)  + 2 c_{\rm G} \bar{f}'_{\chi} \bar{H}^{2} x  
\end{eqnarray} 
and all quantities are in dimensionless form, i.e.\ $\bar{H} = H/H_{0}$, $\bar{f} = f/(M_{\rm pl}^{2}H_{0}^{2})$, and primes denote derivatives with respect to $N=\ln a$. 

In the de Sitter limit for the case $c_1=0$, one finds $\geff/G_N=1/\kappa_1$. However, although both $\chi$ and $x$ approach 0, they do so such that $f_{\chi} \chi \to 0$, $f_{\chi} x^{2} \to {\rm const}$, and $f_{\chi\chi}x^{4} \to \infty$.  Thus $|\kappa_7|\gg|\kappa_{1}| \to +\infty$ and $\geff\to0$. 
That is, gravity appears to turn off at late times.  
This arises in this limit from the nonlinear structure of the theory, 
i.e.\ the presence of $f_{\chi\chi}$ and its power law behavior. 

\begin{figure}[htbp!]
\includegraphics[angle=-90,width=0.49\textwidth]{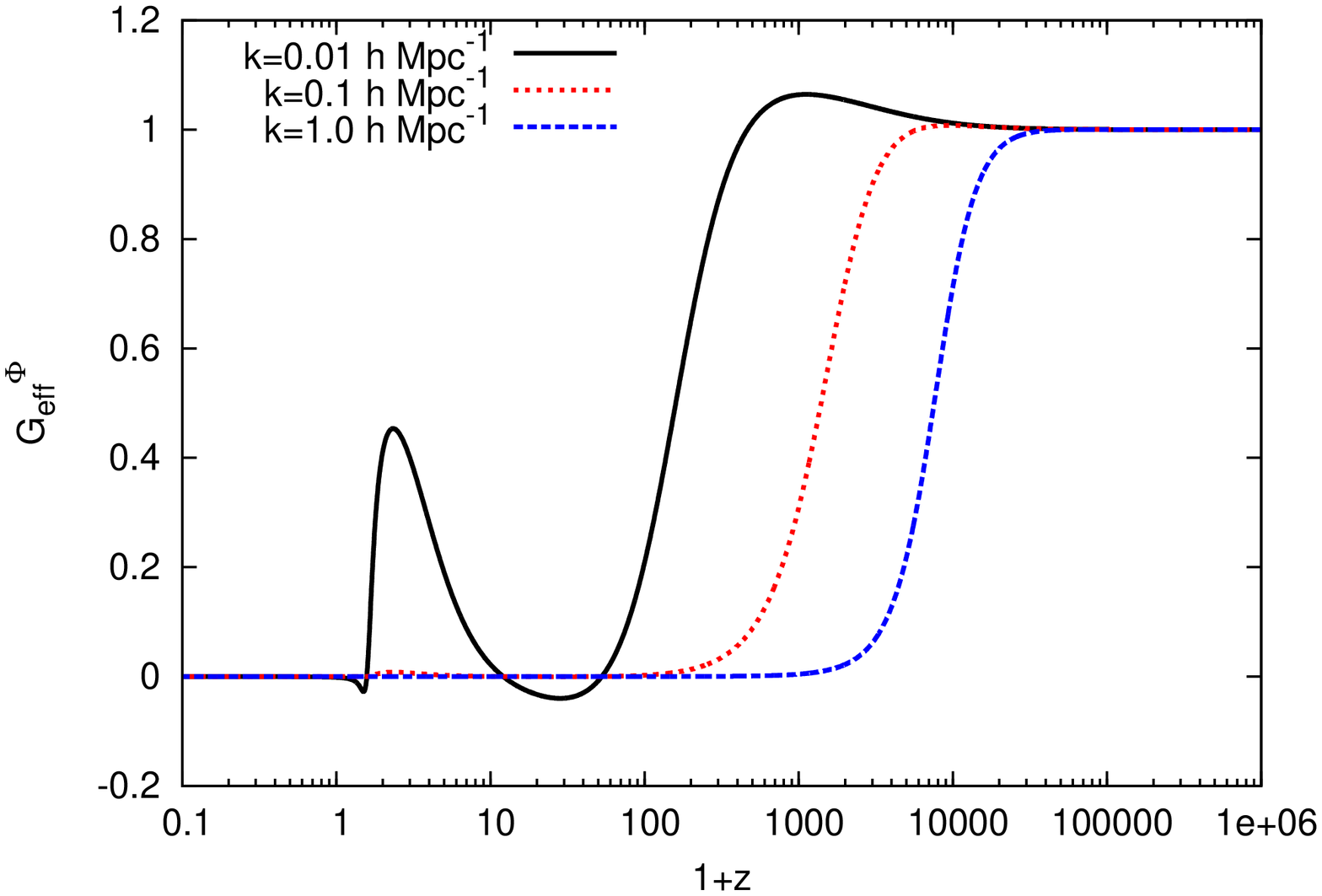}
\includegraphics[angle=-90,width=0.49\textwidth]{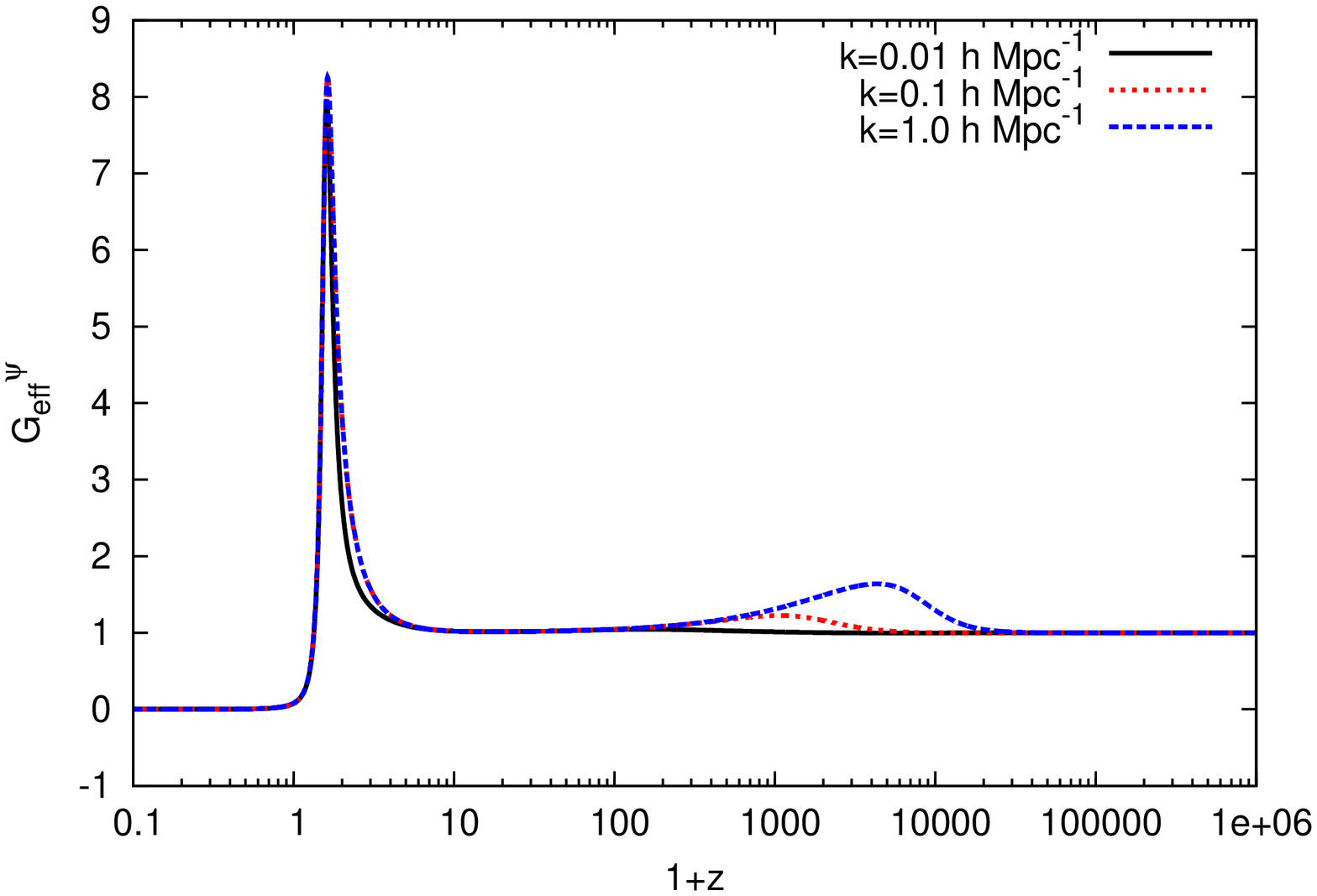} 
\caption{The gravitational couplings $G_{\rm eff}$ in the nonlinear theory 
(left panel $G_{\rm eff}^{(\Phi)}$, right panel $G_{\rm eff}^{(\psi)}$, 
both for $f=\chi^{0.9}$) become scale dependent, as shown by the 
evolution for three different wave modes: $k=0.01\,h/{\rm Mpc}$ (solid black), 
$k=0.1\,h/{\rm Mpc}$ (dotted red) and $k=1.0\,h/{\rm Mpc}$ (dashed blue). 
Gravity vanishes at late times.} 
\label{fig:geffs}
\end{figure}

The numerical solutions for the evolution $\geff(z)$ are shown in 
Fig.~\ref{fig:geffs}.  At high redshift the 
theory acts as general relativity, then deviations begin when 
$(k/aH)^2 \Omega_\phi^{2}\sim 1$.  At this point, the $\kappa_{7}$ 
contribution to 
$\geff$ will dominate due to the $k^2$ term. Since $\kappa_{7}$ appears in 
the denominator of $\geff^{(\Phi)}$, this scale dependent 
effective Newton's constant will 
typically vanish at high redshift, during matter domination
(the exact redshift will be scale dependent and will also be determined by the 
initial conditions for the scalar field energy density). $\geff^{(\psi)}$ on the other
hand will not vanish at early times owing to its different $\kappa_{7}$ 
dependence.  

The gravitational coupling $\geff^{(\psi)}$ entering matter growth 
behaves as GR until near the present, since large $\kappa_{7}$ actually 
cancels out from it.  At low redshift it spikes and then 
vanishes.  The gravitational coupling $\geff^{(\Phi+\psi)}$ entering 
light deflection is given by the mean $[\geff^{(\Phi)}+\geff^{(\psi)}]/2$ 
and so shows deviations at both high and low redshift.  We emphasize 
that the current model is not proposed as observationally viable but rather to 
introduce interesting theoretical properties of nonlinear, noncanonical 
kinetic gravity. 

\subsection{Ghost and Stability Conditions} \label{sec:ghost} 

In order to find the ghost conditions, we need the action for the independent degrees of freedom. It is convenient for this task to evaluate Eq.~(\ref{eq:actio2}) in the flat gauge (i.e.\ $\Phi=0$). Then we can see that the fields $\psi,\beta$ and $\delta\chi$ can be integrated out leaving only two scalars to propagate, i.e.\ $\delta\phi$ (the new-gravity mode), and $v$ (the matter mode). 
But there is a crucial subtlety: the
quadratic term $\delta\chi^{2}$ will generate a term proportional
to $k^{4}\delta\phi^{2}/a^{4}$. This means that this theory will
modify the high $k$ behaviour of the modes, and this will lead to
possible cosmological signatures. This situation is similar to what
happens for FLRW backgrounds in the $f(R,G_{\rm GB})$ theories 
\cite{frgb}. This
$k^{4}$-dependent term vanishes when $f_{\chi\chi}=0$,
that is when the action is linear in the combination 
$c_{2}X+(c_{G}/M^2)G^{\mu\nu}\phi_{\mu}\phi_{\nu}$.  
This behaviour is a typical signature of the presence of a massive
mode ($\delta\chi$), whose kinetic term vanishes, but not its mass.

After removing the auxiliary field $\delta\chi$, and this is possible
only when $f_{\chi\chi}\neq0$, we can write down the action as 
\begin{equation}
\label{eq:acc} S=\int d^{4}xa^{3}\!\left[A_{ab}\dot{V}_{a}\dot{V}_{b}+\frac{B\epsilon_{ab}}{a^{2}}\,(\partial_{i}\dot{V}_{a})(\partial_{i}V_{b})-\frac{D_{ab}}{a^{4}}\,(\partial^{2}V_{a})\,(\partial^{2}V_{b})-\frac{E_{ab}}{a^{2}}\,(\partial_{i}V_{a})(\partial_{i}V_{b})+C\epsilon_{ab}\dot{V}_{a}V_{b}+M_{ab}V_{a}V_{b}\right],
\end{equation}
where we have defined $V_{1}=\delta\phi$, $V_{2}=v$. The matrices
$A,D,E,M$, as well as the two coefficients $B$ and $C$, are functions 
of the background.  
Here we have also defined $\epsilon_{ab}$ as the two
dimensional antisymmetric matrix with $\epsilon_{12}=1$. Furthermore,
the only non-zero matrix element of the matrix $D$ corresponds to
$D_{11}$.

The no-ghost requirements are
\begin{eqnarray} 
\nonumber & & {\rm det [A]} = \left[ f_{\chi\chi} \left\{ c_{\rm G} \left( c_{2} + 6c_{\rm G} \bar{H}^{2}\right)\left(12c_{\rm G}\bar{H}^{2} - c_{2} \right) \bar{H}^{4} x^{4} f_{\chi} + \bar{H}^{2}x^{2} \left( c_{2}^{2} - 6c_{1}c_{\rm G}^{2}\bar{H}^{4}x^{2} + 36 c_{\rm G}^{2} \bar{H}^{4} + 12 c_{2} c_{\rm G} \bar{H}^{2}\right)\right\} \right. \\ 
& & \qquad \left. +c_{\rm G} \left(18c_{\rm G} \bar{H}^{2} - c_{2}\right) \bar{H}^{2}x^{2}f_{\chi}^{2} + \left(c_{2} + 6c_{\rm G} \bar{H}^{2} - c_{1}c_{\rm G}\bar{H}^{2}x^{2} \right) f_{\chi} + c_{1} \right] 
{(1+w)\bar{\rho}_{\rm w} \left( 1 - c_{\rm G} \bar{H}^{2}x^{2} f_{\chi}\right) \over 4\Delta_{2}} > 0 \\ 
\nonumber & & A_{22} = \left[ \left( c_{\rm G}^{2} \left(36 c_{\rm G}\bar{H}^{2} + 5c_{2}\right) f_{\chi}\bar{H}^{6}x^{6} - c_{\rm G} \bar{H}^{4}x^{4}\left(c_{1}c_{\rm G}\bar{H}^{2}x^{2} + 18c_{\rm G} \bar{H}^{2} + 2c_{2}\right)\right)f_{\chi\chi} + 9c_{\rm G}^{2} \bar{H}^{4}x^{4} f_{\chi}^{2} - 6 c_{\rm G} \bar{H}^{2}x^{2}f_{\chi} + 1 \right] \\ 
\label{eq:a1} & & \qquad\times {(1+w)\bar{\rho}_{\rm w} \over 2\Delta_{2}}>0 
\end{eqnarray} 
where in addition to the scalar field we have assumed the presence of a 
barotropic fluid with equation of state $w$ and energy density 
$\bar{\rho}_{\rm w} = \rho_{\rm w}/(H_{0}^{2}M_{\rm pl}^{2})$, and 

\begin{eqnarray} 
\nonumber & &  \Delta_{2} = \left[ \left\{\bar{H}^{6}x^{6} c_{\rm G}^{2}\left(36c_{\rm G}\bar{H}^{2} + 5c_{2}\right)f_{\chi} -  c_{\rm G}^{2}\bar{\rho}_{\rm w} \bar{H}^{4}x^{4} - c_{\rm G}\bar{H}^{4}x^{4} \left(c_{\rm G}c_{1}\bar{H}^{2}x^{2} + 18c_{\rm G}\bar{H}^{2} + 2c_{2}\right)\right\} f_{\chi\chi} \right. \\   
& \qquad & \left. + 9c_{\rm G}^{2}\bar{H}^{4}x^{4}f_{\chi}^{2} - 6c_{\rm G} \bar{H}^{2}x^{2}f_{\chi} + 1 \right] w - c_{\rm G}^{2} \bar{\rho}_{\rm w} \bar{H}^{4}x^{4} f_{\chi\chi} \ . 
\end{eqnarray}

During the radiation era, $\Delta_2\approx w=1/3$ and the square 
brackets in $A_{22}$ resolve to 1, so indeed $A_{22}>0$.  
In the matter era where 
$w=0$, then $A_{22}$ has the same sign as $-f_{\chi\chi}$ and so we require 
$n<1$ in the power law model $f\sim \chi^n$.  For $\det A$, since 
$\bar H^2\gg1$ then the $c_G$ terms will dominate in the early universe 
over the other scalar field terms in the absence of fine tuning them to 
be small.  This results in the condition $(2n-1)/(1-n)>0$, satisfied for 
$1/2<n<1$.  Thus such theories are free of ghosts. 

Checking the speed of propagation of the field, with Laplace 
stability given by nonnegative sound speed squared, $c_s^2\ge0$, is 
somewhat more involved. In the high-$k$ limit, we find that the dispersion relations are given by 
\begin{eqnarray}
\omega_{\phi}^{2} & = & \frac{B^{2}+A_{22}D_{11}}{\det[A]}\,\frac{k^{4}}{a^{4}}=\frac{16c_{G}^{4}f_\chi^{2}\bar H^{8}x^{6}f_{\chi\chi}}{(1-c_{G}f_\chi \bar H^2 x^2)\Delta}\,\frac{k^{4}}{a^{4}}\,,\\
\omega_{{\rm pf}}^{2} & = & \frac{D_{11}E_{22}}{B^{2}+A_{22}D_{11}}\,\frac{k^{2}}{a^{2}}=w\,\frac{k^{2}}{a^{2}}\,,
\end{eqnarray}
where $\Delta$ is defined as 
\begin{eqnarray}
\Delta & \equiv & f_{\chi\chi}\,\left[c_{{G}}(c_{{2}}+6\, c_{{G}}\bar H^{2})(c_{{2}}-12\, c_{{G}}{\bar H}^{2})\bar H^4 x^4 f_\chi+6c_{1}\,{c_{{G}}^{2}}\bar H^6 x^4-  (c_{{2}}+6\, c_{{G}}{\bar H}^{2})^{2}\bar H^2 x^2\right]\nonumber \\
 &  & {}-c_{{G}}(18\, c_{{G}}{\bar H}^{2}-c_{{2}})\bar H^2 x^2 f_\chi^{2}+[c_{1}c_{{G}}\bar H^2 x^2- (c_{{2}}+6\, c_{{G}}{\bar H}^{2})]f_\chi-c_{1} \,.
\end{eqnarray} 

The speeds of propagation are then found as the group velocity $c=a\partial\omega/\partial k$,
or 
\begin{equation}
\label{eq:aa1} c_{\phi}^{2}=\frac{64c_{G}^{4}f_\chi^{2}\bar H^{8} x^6 f_{\chi\chi}}{(1-c_{G}f_\chi \bar H^2 x^2)\Delta}\,\frac{k^{2}}{a^{2}}\geq0\,,\qquad c_{{\rm pf}}^{2}=w\geq0\,.
\end{equation}
One of the two solutions is trivial as it corresponds to the speed
of the perfect fluid, but the other one sets a stability condition,
and states that the speed of propagation will be scale dependent.

Because of the $k^4$ terms in Eq.~(\ref{eq:acc}), the dispersion relation 
and hence sound speed will be wavenumber dependent.  From 
Eq.~($\ref{eq:aa1}$), we see that in the high $k$ limit the leading order 
contribution to $c_{\phi}^{2}$ goes like 
\begin{equation} 
c_{\phi}^{2} \sim \left( {k \over aH}\right)^{2} \Omega_{\phi}^{2} 
\end{equation} 
where $\Omega_{\phi} = 8\pi G \rho_{\phi}/(3H^{2})$. During matter domination, we typically find
$k/(aH) \gg 1$ for sub-horizon modes relevant to linear perturbation theory, and $\Omega_{\phi} \sim {\cal O} \left(10^{-2} \Omega_m \right) $ (this is largely dependent upon the initial conditions imposed, however this is a conservative upper bound on how large
$\Omega_{\phi}$ can be during matter domination). Hence we expect the $k^{4}$ 
term in Eq.~(\ref{eq:acc}) to be the dominant contribution to $c_{\phi}^{2}$ between $z\sim (1,1000)$. 

If this held for radiation domination, then 
$c_s^2 \sim (k/aH)^2   \Omega_{\phi}^{2} (1-n)/(2n-1)$ 
and so the theory would be 
Laplace stable for $1/2<n<1$.  However, in the linear case $n=1$, the 
terms proportional to $f_{\chi\chi}$ vanish identically and the leading 
order $k^2$ contribution vanishes, leaving a scale independent sound 
speed.  As found in \cite{applin}, the linear theory composed of a standard 
kinetic term and a kinetic term coupled to the Einstein tensor (i.e.\ the 
purely kinetic gravity theory of \cite{gub} with the canonical kinetic 
term generalized to have an arbitrary constant coefficient) is Laplace 
unstable in the radiation era.  Thus, the nonlinearity of the current model 
can avoid that instability. 

However, at early enough times during radiation domination $\Omega_\phi$ 
drops so low that the $(k/aH)^2\Omega_\phi^2$ term becomes subdominant for 
the modes $k$ relevant to linear perturbation theory.  To calculate the 
leading order behaviour of $c_{\phi}$ in the very early Universe, it is 
more instructive to consider the issue from a different angle. 

To preserve the CMB acoustic peak structure, and also to obtain an expansion history consistent with observations, we want initial conditions during radiation domination such that the effect of the scalar field $\phi$ on the background expansion and the metric perturbations is negligible. In this case we can assume that $\Omega_{\phi} \ll 1 $ and the metric potentials are sourced by density perturbations only, hence we can treat the scalar field perturbations $\delta \phi$ as evolving on an otherwise standard cosmological background. Under this assumption, one can analytically calculate the no-ghost and Laplace conditions. For the linear model $f(\chi) \sim \chi$, it was found in \cite{applin} that the scalar field perturbations possessed a sound speed $c_{\phi}^{2}$ that was negative during radiation domination; $c_{\phi}^{2} = -1/3$. This would lead to exponential growth of the scalar field perturbations, destroying the standard cosmological picture. One can perform a similar analysis for the more general $f(\chi)$ case. We find the following no-ghost and Laplace stability conditions 
\begin{equation}  
3 c_{\rm G} \bar{H}^{2} \left( f_{\chi} + 2\chi f_{\chi\chi}\right) > 0 
\end{equation} 
and 
\begin{equation} 
{2 \dot{H} + 3H^{2} \over 3H^{2}} {f_{\chi} \over f_{\chi} + 2\chi f_{\chi\chi}} > 0 
\end{equation} 
where we assume that at early times the $c_{\rm G}$ contribution to the 
$\phi$ energy density is much larger than the standard canonical $c_{2}$ term (which is valid barring an extreme fine tuning of $c_{\rm G}$). For the power law models, during radiation domination these conditions correspond to 
\begin{eqnarray} 
c_{\rm G}  f_{\chi} &<& 0 \\ 
2n-1 &<& 0 \ , 
\end{eqnarray} 
which would violate the positivity of $\rho_{\phi}$.  This could potentially cause problems in the transition to the matter dominated era, 
where $\rho_{\phi} > 0$ is required to ensure that the no-ghost condition is satisfied. 
We conclude that the power law models cannot simultaneously satisfy the 
no-ghost and Laplace stability requirements at all times while also having 
$\rho_{\phi} > 0$ during radiation domination. This does not preclude the 
possibility that a non-power law model might be constructed that can. 

The presence of a scale-dependent speed of propagation is a feature of this model and it has physical implications, especially at late times when $\Omega_\phi\to1$. The reason for the presence of such a term may be due to the large symmetries of the FLRW manifolds, similarly to what happens in the context of $f(R,G_{\rm GB})$ theory (see e.g.\ \cite{frgb}). In that case it was shown that the kinetic term of one of the scalar perturbation modes was vanishing on general FLRW manifolds, so that it could be integrated out from the Lagrangian, giving rise in this way to a scale dependent speed of propagation $c_s^2\propto k^2/a^2$.  
Furthermore, also in that theory, at late times as the $k^2$-regime starts dominating, gravity for high $k$'s tends to become weaker and weaker, i.e.\ $G_{\rm eff}/G_N\to0$. It would be interesting to study, along the same lines of the $f(R,G_{\rm GB})$ theories, whether anisotropic models (such as Bianchi-I type manifolds) possess more propagating  degrees of freedom than FLRW. 
This will be investigated in a future project. Nonetheless, if indeed this scenario does happen, then this theory would behave similarly also to massive gravity, as it was shown that for that theory the kinetic terms of three perturbation modes vanish on FLRW due to the high degree of symmetries of the background \cite{massiveG}.

\section{Conclusions} \label{sec:concl} 

Gravitation is a fundamental force that we have just begun to explore 
cosmologically.  One of the great advances made in gravity research in 
the past few years is the realization that symmetry principles both 
strongly restrict the theory and open up new avenues and effects.  Galileon 
gravity and massive gravity both use shift symmetric fields and their 
couplings to functions of the metric to enable new properties, including 
cosmic acceleration without a cosmological constant or field potential. 
An action allowed by the symmetries and well behaved in initial value 
formulation, specifically one leading to second order equations of motion, 
is of particular interest.  If moreover the field exhibits self tuning, 
allowing it to overcome a high energy cosmological constant, the theory 
is well worth examining. 

We show that by promoting a purely kinetic gravity term to a nonlinear 
function, possibly mixed with a noncanonical kinetic term for the field, 
fascinating properties can ensue.  In addition to second order equations 
of motion and self tuning, the theory does not incur extra propagating 
degrees of freedom on a cosmological background.  
Similar effects of symmetric backgrounds are seen in massive gravity.  
For example, in massive gravity it was found \cite{massiveG} that in 
isotropic spacetimes the shift symmetric (St\"uckelberg) fields have 
vanishing mixing between the graviton and the scalar mode, and furthermore 
their kinetic terms vanish. 

The new term discussed here, ``Fab 5 Freddy,'' can self accelerate and is 
merely the harbinger of a whole 
class of such nonlinear promotions or combinations of terms. 

The background evolution of the expansion and field lead to early time 
tracker behavior and late time de Sitter attractors.  Solving the linear 
perturbation equations we see that simple power law functions can be free 
of ghosts.  The gravitational coupling and dispersion relation of 
perturbations become scale dependent, possibly leading to an early time 
instability and a late time vanishing of gravity.  The specific models 
studied may not be observationally viable but the characteristics arising 
from the nonlinear, noncanonical action open new aspects of gravity.  
Most intriguing is the self tuning property that can cancel a bare 
cosmological constant dynamically, even through phase transitions.  
The evolution of the field basically makes $\Lambda$ invisible. 

That the most general scalar-tensor theory giving second order field 
equations in 4D could be further generalized, at least on cosmological 
backgrounds, is highly interesting. 
The specific term considered here, a nonlinear promotion of the field 
kinetics coupled to the Einstein tensor (the unique, low mass dimension 
shift symmetric combination giving second order field equations), is 
merely a proof of principle, while theoretically instructive.  Ways to 
extend this class of theory more generally, to different nonlinear 
functions and combinations, are straightforward and may preserve the most 
interesting and desirable characteristics while leading to more 
viable predictions experimentally.

\acknowledgments 

ADF is grateful for the warm hospitality provided at the Institute for 
the Early Universe where the project was initiated.  
This work has been supported by World Class University grant 
R32-2009-000-10130-0 through the National Research Foundation, Ministry 
of Education, Science and Technology of Korea, and in part by the Director, 
Office of Science, Office of High Energy Physics, of the 
U.S.\ Department of Energy under Contract No.\ DE-AC02-05CH11231.

\appendix 

\section{Covariant Equations of Motion} \label{sec:coveom} 

The scalar field equation is given by 
\begin{equation} 
c_{1} \Box \phi + c_{2} \nabla_{\alpha} \left[ f_{\chi} \nabla^{\alpha} \phi \right] - 2 c_{\rm G} G^{\alpha\beta} \nabla_{\alpha}\left[ f_{\chi} \nabla_{\beta}\phi \right]  = 0 \ . 
\end{equation} 
The $\chi$ field is given by 
\begin{equation} 
\chi = - {c_{2} \over 2} \nabla_{\alpha}\phi \nabla^{\alpha}\phi + c_{\rm G} G^{\alpha\beta}\nabla_{\alpha}\phi \nabla_{\beta}\phi   \ . 
\end{equation} 
The Einstein equations are given by 
\begin{equation} 
G_{\mu\nu} = 8\pi G \left[ T_{\mu\nu}^{({\rm mat})}+T_{\mu\nu}^{({\rm rad})}+T_{\mu\nu}^{(\phi)} \right] \ , 
\end{equation}
where 
\begin{eqnarray}  
\nonumber & & T_{\mu\nu}^{(\phi)} =    - c_{\rm G} f_{\chi} \left[ g_{\mu\nu} \Box\phi \Box\phi - 2 \Box\phi \nabla_{\mu}\nabla_{\nu}\phi + 2\nabla_{\mu}\nabla_{\lambda}\phi \nabla_{\nu}\nabla^{\lambda}\phi - g_{\mu\nu} \nabla_{\lambda}\nabla_{\alpha}\phi \nabla^{\lambda}\nabla^{\alpha}\phi \right] \\ \nonumber & & \qquad \qquad \qquad + c_{\rm G} f_{\chi} \left[ R_{\mu\nu}\nabla_{\alpha}\phi \nabla^{\alpha}\phi + R \nabla_{\mu}\phi \nabla_{\nu}\phi - {1 \over 2} g_{\mu\nu} R \nabla_{\alpha}\phi \nabla^{\alpha}\phi \right]  \\ \nonumber & & \qquad \qquad \qquad   - 2 c_{\rm G} f_{\chi} \left[ R_{\lambda \nu}\nabla^{\lambda}\phi \nabla_{\mu} \phi  + R_{\lambda\mu} \nabla^{\lambda}\phi \nabla_{\nu}\phi - g_{\mu\nu} R_{\rho\lambda}\nabla^{\rho}\phi \nabla^{\lambda}\phi + R^{\sigma}{}_{\mu\beta\nu} \nabla^{\beta}\phi \nabla_{\sigma}\phi \right]   \\ \nonumber & & + 2c_{\rm G} \left[  (\nabla_{\alpha}\nabla_{(\mu}f_{\chi})\nabla_{\nu)}\phi  \nabla^{\alpha}\phi - {1 \over 2} (\Box f_{\chi}) \nabla_{\mu}\phi\nabla_{\nu}\phi - {1 \over 2} g_{\mu\nu} (\nabla_{\alpha}\nabla_{\beta}f_{\chi}) \nabla^{\alpha}\phi \nabla^{\beta}\phi -{1 \over 2}(\nabla_{\mu}\nabla_{\nu}f_{\chi})\nabla_{\alpha}\phi\nabla^{\alpha}\phi + {1 \over 2} g_{\mu\nu} (\Box f_{\chi})\nabla_{\alpha}\phi \nabla^{\alpha}\phi   \right]  \\ \nonumber & & + 2c_{\rm G}  \left[   \nabla_{(\mu} f_{\chi} \nabla_{\nu)}\phi \Box \phi - \nabla_{(\mu} f_{\chi} \nabla_{\nu)}\nabla_{\alpha} \phi \nabla^{\alpha}\phi - \nabla_{\alpha} f_{\chi} \nabla^{\alpha}\nabla_{(\mu}\phi \nabla_{\nu)}\phi -  g_{\mu\nu} \nabla^{\alpha}f_{\chi} \nabla_{\alpha}\phi \Box \phi + g_{\mu\nu} \nabla^{\alpha}f_{\chi} \nabla^{\beta}\phi \nabla_{\alpha}\nabla_{\beta}\phi \right. \\ & & \left. + \nabla_{\alpha}f_{\chi}\nabla^{\alpha}\phi \nabla_{\mu}\nabla_{\nu}\phi    \right]   +  g_{\mu\nu}f -  g_{\mu\nu} f_{\chi} \chi   + c_{1} \left[ \nabla_{\mu}\phi \nabla_{\nu}\phi - {1 \over 2} g_{\mu\nu} \nabla_{\alpha}\phi \nabla^{\alpha}\phi \right]   
+ c_{2} f_{\chi} \left[ \nabla_{\mu}\phi \nabla_{\nu}\phi - {1 \over 2} g_{\mu\nu} \nabla_{\alpha}\phi \nabla^{\alpha}\phi \right]  ,
\end{eqnarray} 
and parentheses in a subscript denote symmetrization of the indices.

\section{Perturbation Equations in Detail} \label{sec:apxpert} 

\subsection{The scalar perturbations}
Let us write down the perturbed metric in the following form
\begin{equation}
ds^2=-(1+2\psi) dt^2 +2\partial_i \beta dt\,dx^i + a^2\,(1-2\Phi) d{\bm x}^2\,.
\end{equation}
Expanding the scalar field as $\phi=\phi(t)+\delta\phi$, and considering 
a barotropic perfect fluid with equation of state $P=w\rho$ (for an action 
approach of perfect fluids see e.g.\ \cite{perfF}), 
then we find that in Fourier space, the action at second order in the perturbation fields can be written as 
\begin{eqnarray}
S^{(2)}&=&\int dt d^3{x}\, {a}^{3} \left\{ 
- \left( W_{{1}}\psi +W_{{2}}\dot{\delta\phi}-W_{{3}}\dot{\Phi}-W_{{4}}\delta \phi -\rho\left( 1+w \right) V+W_{{5}}\delta \chi  \right) \frac{\partial^{2}\beta}{{a}^{2}}+ 
\frac12\left( {\frac {\rho\, \left( 1+w \right) }{w}}-W_{{6}} \right) {\psi}^{2}\right.\nonumber\\
&-& \left( W_{{7}}\dot{\delta\phi}+W_{{8}}\dot{\Phi}+{\frac {\rho\, \left( 1+w \right)  \left( \dot{V}-3\,wHV \right) }{w}}-W_9{\frac {\partial^2\delta \phi }{{a}^{2}}}+W_{10}{\frac {\partial^2 \Phi }{{a}^{2}}}+W_{{11}}\delta \chi  \right)  \psi
+\frac12\,W_{{12}}{\dot{\delta\phi}}^{2}
+\frac12\,W_{{13}}{\dot{\Phi}}^{2}\nonumber\\
&-&\frac12\,W_{{14}}\dot{\Phi}\dot{\delta\phi}
-\frac12\,W_{15}\,{\frac {(\partial\delta \phi) ^{2}}{{a}^{2}}}
-\frac12\,W_{16}\,{\frac {(\partial{\Phi})^{2}}{{a}^{2}}}
+\frac12\,{\frac {\rho\, \left( 1+w \right) {\dot{V}}^{2}}{w}}
-\frac12\,{\frac {\rho\, \left( 1+w \right) {k}^{2}{V}^{2}}{{a}^{2}}}
-\frac12\,f_{{\chi \chi }}{\delta \chi }^{2}\nonumber\\
&+&\left. \left( W_{{17}}\dot{\delta\phi}-W_{{18}}\dot{\Phi}
+W_{19}{\frac {\partial^2\Phi }{{a}^{2}}} \right) \delta \chi 
- \left( -W_{20}\frac {\partial^2\delta \phi }{{a}^{2}}-9\,\rho\,wH \left( 1+w \right) V+3\, \left( 1+w \right) \rho\,\dot{V}+W_{{21}}\dot{\delta\phi} \right)  \Phi  \right\} ,\label{eq:actio2}
\end{eqnarray}
where the matter field $V$ is the scalar component of $\delta T^0{}_i=-\rho(1+w)\partial_i V$, so that the matter density contrast $\delta_m=\delta\rho/\rho$ can be written as $w\delta_m/(1+w)=\dot{V}-3wHV-\psi$. 

Notice we still have one gauge degree of freedom to choose. For example, we can consistently set $\beta=0$ (Newtonian gauge), or $\delta\phi=0$ (uniform field gauge), or $\Phi=0$ (flat gauge).

The coefficients of the previous action are the following: 
\begin{eqnarray}
W_1&=&2\,H\Mpl^{2}-6\,H{c_{{G}} \over M^{2}}f_{{\chi}}{\dot\phi}^{2}\,,\\
W_2&=&W_9=4\, {c_{{G}} \over M^{2}}f_{{\chi}}H{\dot\phi}\,\\ 
W_3&=&W_{10}=-2\,\Mpl^{2}+2\,{c_{G} \over M^{2}}f_{{\chi}}{\dot\phi}^{2}\,,\\
W_4&=&{\dot\phi}c_2\,f_{{\chi}}+{\dot\phi}c_1+6\,{\dot\phi}{c_{G} \over M^{2}}
{H}^{2}f_{{\chi}}\,,\\
W_5&=&2\,{c_{G} \over M^{2}}f_{{\chi \chi }}H{\dot\phi}^{2}\,,\\
W_6&=&-c_{{1}}{\dot\phi}^{2}+6\,\Mpl^{2}{H}^{2}-36\,{c_{G} \over M^{2}}f_{{\chi}
}{H}^{2}{\dot\phi}^{2}-c_{{2}}f_{{\chi}}{\dot\phi}^{2}\,,\\
W_7&=&18\,{\dot\phi}{c_{G} \over M^{2}}{H}^{2}f_{{\chi}}+{\dot\phi}c_{{1}}+{\dot\phi}c_{{2}}f_{{\chi}}\,,\\ 
W_8&=&6\,H\Mpl^{2}-18\,H{c_{G} \over M^{2}}f_{{\chi}}{\dot\phi}^{2}\,,\\
W_{11}&=&f_{{\chi \chi }}{\dot\phi}^{2}c_{{2}}+12\,f_{{\chi \chi }}{\phi_{{1}
}}^{2}{c_{G} \over M^{2}}{H}^{2}\,,\\
W_{12}&=&6\,{c_{G} \over M^{2}}f_{{\chi}}{H}^{2}+c_{{1}}+c_{{2}}f_{{\chi}}\,,\\
W_{13}&=&-6\,\Mpl^{2}+6\,{c_{G} \over M^{2}}f_{{\chi}}{\dot\phi}^{2}\,,\\
W_{14}&=&24\,{c_{G} \over M^{2}}f_{{\chi}}H{\dot\phi}\,,\\
W_{15}&=&c_{{2}}f_{{\chi}}+4\,{c_{G} \over M^{2}}f_{{\chi}}\dot H+c_{{1}}+6\,c_{{
G}}f_{{\chi}}{H}^{2}\,,\\
W_{16}&=&-2\,{c_{G} \over M^{2}}f_{{\chi}}{\dot\phi}^{2}-2\,\Mpl^{2}\,,\\
W_{17}&=&{\dot\phi}f_{{\chi \chi }}c_{{2}}+6\,{\dot\phi}f_{{\chi \chi }}{c_{G} \over M^{2}}{H}^{2}\,,\\
W_{18}&=&6\,{c_{G} \over M^{2}}f_{{\chi \chi }}H{\dot\phi}^{2}\,,\\
W_{19}&=&2\,f_{{\chi \chi }}{\dot\phi}^{2}{c_{G} \over M^{2}}\,,\\
W_{20}&=& \left[ -4\,{c_{G} \over M^{2}}f_{{\chi}}-4\,{c_{G} \over M^{2}}{\dot\phi}^{2} \left( 6\,c_{
{G}}{H}^{2}+c_{{2}} \right) f_{{\chi \chi }} \right] \, {\ddot\phi}-4\,c_{
{G}}f_{{\chi}}H{\dot\phi}-24\,{{c_{G} \over M^{2}}}^{2}f_{{\chi \chi }}{{\dot\phi}
}^{3}H \dot{H}\,,\\
W_{21}&=&3\,{\dot\phi} \left( 6\,{c_{G} \over M^{2}}f_{{\chi}}{H}^{2}+c_{{1}}+c_{{2}}f_{{\chi}} \right).
\end{eqnarray}
The equations of motion for the perturbations in any gauge can be derived by using standard variational calculus.

\subsection{Tensor perturbations}
By introducing the two polarizations of transverse and traceless perturbations 
for the metric, we can write down the action expanded at second order as
\begin{equation}
S^{(2)}_{\rm GW}=\sum_{\lambda=+,\times}\int dt d^3{x}\, a^3\left[\frac18\left(\Mpl^2-{c_G \over M^{2}} f_{\chi}\dot\phi^2\right)\,\dot{h}_\lambda^2-\frac18\left(\Mpl^2+ {c_G \over M^{2}} f_{\chi}\dot\phi^2\right)\frac{(\partial h_\lambda)^2}{a^2}\right] .
\end{equation}
This gives the no-ghost condition $1-c_G f_{\chi}\bar{H}^{2}x^{2}>0$, and speed of propagation equal to 
\begin{equation}
c_{\rm GW}^2=\frac{1+c_G f_{\chi}\bar{H}^{2}x^{2}}{1-c_G f_{\chi}\bar{H}^{2}x^{2}}\,.
\end{equation} 
Note that due to the coupling to the Einstein tensor this is not equal 
to the speed of light.  A stable evolution for the background requires 
that $c_{\rm GW}^2\geq0$.


\end{document}